\unless\ifcsname FFFvariant\endcsname
\fi
\documentclass[graybox, secnum]{svmult}

\usepackage{newtxtext,newtxmath} 
\usepackage{helvet}         
\usepackage{courier}        
\usepackage{type1cm}        
\usepackage{mathrsfs}
\usepackage{makeidx}         
\usepackage{graphicx}        
\usepackage{multicol}        
\usepackage[bottom]{footmisc}
\usepackage{hyperref}        
\usepackage{soul}            
\hypersetup{colorlinks=true,urlcolor=blue,citecolor=orange}
\usepackage[square,numbers]{natbib}
\makeindex             
\usepackage[textsize=tiny,backgroundcolor=yellow]{todonotes}
\usepackage{dsfont}%
\usepackage{booktabs}%
\usepackage{cancel}%
\usepackage{subcaption}%
\usepackage[inline]{enumitem}%
\usepackage{mathtools}
\usepackage{cleveref}
\usepackage[nolist]{acronym}
\usepackage{nicefrac}
\usepackage{xspace}
\usepackage{mleftright}
\mleftright
\makeatletter
\g@addto@macro\bfseries{\boldmath}
\makeatother
\newlist{goals}{enumerate*}{1}
\setlist[goals]{label=(\arabic*),ref=(\arabic*)}
\crefname{goalsi}{goal}{goals}
\Crefname{goalsi}{Goal}{Goals}
\newlist{challenges}{enumerate}{1}
\setlist[challenges]{label=\arabic*.,ref=\arabic*}
\crefname{challengesi}{challenge}{challenges}
\Crefname{challengesi}{Challenge}{Challenges}

\newcommand{\dd}{\mathop{}\!\mathrm{d}}
\newcommand{\ii}{\hskip0.1ex\mathrm{i}\hskip0.1ex}
\newcommand{\ee}{\mathrm{e}}

\DeclareMathOperator{\lcm}{lcm}
\newcommand{\cc}[2]{c\genfrac{[}{]}{0pt}{}{#1}{#2}}
\newcommand{\one}[0]{\mathds{1}}
\newcommand{\Bsquare}{{\Large$\Box$}}
\newif\ifBookChapter

\begin{document}
\title*{Free Fermionic Constructions of Heterotic Strings}
\author{Ioannis Florakis  and
	John Rizos}%
\institute{
Ioannis Florakis \at Department of Physics,
University of Ioannina, GR45110, Ioannina, Greece, \email{iflorakis@uoi.gr}
\and
John Rizos \at Department of Physics,
University of Ioannina, GR45110, Ioannina, Greece, \email{irizos@uoi.gr}
}
\maketitle 
\abstract{%
This chapter is an introduction to the Free Fermionic Formulation of String Theory, with emphasis on heterotic model building.
After a brief review of bosonization in two dimensional conformal field theories, we discuss how internal bosonic string coordinates can be consistently replaced by free fermionic degrees of freedom. In this framework, worldsheet supersymmetry may be realized entirely among free fermions. Embedding this construction into string theory leads to a number of constraints arising from modular invariance at one and higher genera. The solution of these constraints takes the form of a small number of model building rules from which the string spectrum and interactions may be analyzed. We review some of the most well-studied models in the literature and their classification, with emphasis on the symmetric basis. The explicit map of free fermionic models to the orbifold construction is presented in some detail.
}

\section*{Keywords}
\textcolor{gray}{String Theory, Free Fermionic Construction, Model Building, Heterotic Strings}

\clearpage
\tableofcontents
\markboth{Model Building in Heterotic String}{Model Building in Heterotic String}

\acresetall


\section{Introduction}
\label{sec:FFF_Introduction}

In this chapter we introduce and review the \ac{FFF} of heterotic string theory, together with a discussion of some of its main applications in the context model building.
In the \ac{FFF}, all worldsheet bosonic coordinates are consistently fermionized, except for the non-compactified space-time degrees of freedom
\cite{Antoniadis:1987wp,Antoniadis:1986rn,Kawai:1986va,Kawai:1986ah}. World-sheet 
supersymmetry is preserved although non-linearly realized among the two-dimensional fermionic fields \cite{Antoniadis:1985az}. Factorization of string amplitudes together with modular invariance constraints at one and higher genera, can be explicitly solved and  give rise to a relatively simple set of rules, that allow the direct construction of string models in $d\le 10$ dimensions. Due to exact solvability of the worldsheet \ac{CFT}, the \ac{FFF} machinery allows the complete analysis of the corresponding perturbative string spectra and their interactions. 

Specifically in four space-time dimensions, the \ac{FFF} formalism is particularly suited to the construction of string vacua with rich phenomenological characteristics. 
Well-studied $d=4$ models with interesting phenomenology include the flipped $\text{SU}(5)\times\text{U}(1)$ model \cite{Antoniadis:1989zy}, the Pati--Salam model \cite{Antoniadis:1990hb,Leontaris:1999ce} and the Standard-like Model 
\cite{Faraggi:1989ka}. The classification of 10-dimensional heterotic models in the \ac{FFF} has been presented in \cite{Kawai:1986vd}, while in four dimensions, it has proven very efficient for extensive scans and the classification of huge classes of heterotic string vacua \cite{Faraggi:2004rq,Assel:2010wj,Faraggi:2014hqa,Faraggi:2017cnh}. 

This review is organized as follows. In Section \ref{sec:FFF_fermionization} we begin with a discussion of two-dimensional worldsheet \ac{CFT}'s with free fermionic degrees of freedom and their equivalence to compact bosons at special radii. This equivalence is then demonstrated at the level of the one-loop string partition function. In Section \ref{sec:FFFSUSYModular}, we explain the realization of worldsheet supersymmetry entirely among the free fermionic degrees of freedom, and discuss the structure of the worldsheet supercurrent. Section \ref{sec:FFFSUSYModular} is devoted to the study of the constraints arising from modular invariance at one and higher genera, which leads to the consistent vacuum construction rules. The latter are presented in considerable detail in Section \ref{sec:FFFconstructionRules}, where it is shown that they reduce to a set of basis vectors encoding the fermion boundary conditions, together with a corresponding set of \ac{GGSO}. We illustrate the application of the construction rules in the case of a simple model, by deriving its massless spectrum and superpotential. In Section \ref{sec:FFFSUSYModels} we review some fo the most studied models with $\mathcal{N}=1$ supersymmetry constructed in the \ac{FFF}, including the flipped SU(5), Pati-Salam and Standard Model vacua, and discuss some of their phenomenological properties. Subsequently, in Section \ref{FFF:subsII}, focusing on a symmetric basis that has been much interest in the literature, we discuss the application of the \ac{FFF} to the classification of huge sets of supersymmetric string vacua. Next, in Section \ref{sec:NonSUSY_FFF_Models} we briefly review the construction of non-supersymmetric heterotic models in the context of the \ac{FFF} and its relation to the stringy Scherk-Schwarz mechanism. Finally, in Section \ref{sec:FFForbifold} we present the map between a class of \ac{FFF} models and their equivalent $\mathbb{Z}_2\times\mathbb{Z}_2$ orbifold counterparts, which is particularly useful for deforming the theory away from the fermionic point and addressing questions of supersymmetry breaking.


\section{Fermionization in two dimensions}
\label{sec:FFF_fermionization}

The main idea behind the \ac{FFF} is to exploit the bosonization property \cite{Coleman:1974bu,Mandelstam:1975hb,Witten:1983ar} of chiral fermions in $1+1$ dimensions in reverse, and consistently replace the \ac{CFT} of bosonic coordinates of String Theory ascribed to the internal space with a system of auxiliary free fermions.
In the two-dimensional worldsheet \ac{CFT} with Minkowski signature, one may simultaneously impose both Majorana and Weyl conditions on massless fermions to reduce them to real, one-component spinors. Consider the \ac{CFT} of $2N$ real, non-interacting fermions $\psi^i$, with $i=1,\ldots, 2N$ with action 
\begin{align}
	S = \frac{1}{2\pi}\int \dd^2 z~(\psi^i\bar\partial \psi^i + \bar\psi^i \partial \bar\psi^i) \,.
	\label{FFF:FFaction}
\end{align}
The equations of motion then imply the decomposition into left moving holomorphic spinors $\psi^i(z)$ and right moving anti-holomorphic spinor $\bar\psi(\bar z)$ and reflects the factorization of the free fermion \ac{CFT} into left and right moving sectors, with energy-momentum tensors
\begin{align}
	T(z) = -\frac{1}{2} :\psi^i\partial\psi^i(z): \quad,\quad \bar T(\bar z) = -\frac{1}{2} :\bar\psi^i\bar\partial\psi^i(\bar z): \,.
\end{align}
The conformal anomaly term in the $TT$ and $\bar T\bar T$ \ac{OPE} determines the corresponding central charges of the system as $c_L=c_R=N$.
Similarly, the absence of interacting terms among the $2N$ real fermions leads to the \ac{OPE}s
\begin{align}
	\begin{split}
	&\psi^i(z)\psi^j(w) = \frac{\delta^{ij}}{z-w} +\ldots \,, \\ 
	&\bar\psi^i(\bar z)\bar\psi^j(\bar w) = \frac{\delta^{ij}}{\bar z-\bar w} +\ldots \,,\\
	 &\psi^i(z)\bar\psi^j(\bar z) = {\rm regular}\,. 
	 \end{split}
\end{align}
The action \eqref{FFF:FFaction} enjoys a global ${\rm SO}(2N)_L\times {\rm SO}(2N)_R$ symmetry under $\psi^i \to O_L^{ij}\psi^j$ and $\bar\psi^i \to O_R^{ij}\bar\psi^j$ with $O_L^T O_L = O_R^T O_R = 1$,  generated by the (anti) holomorphic currents
\begin{align}
	J^{ij}(z) = \ii:\psi^i(z)\psi^j(z): \quad,\quad \bar J^{ij}(\bar z) = \ii :\bar\psi^i (\bar z)\bar\psi^j(\bar z): \,,
\end{align}
for $i<j$, which are chirally conserved, $\bar\partial J^{ij}(z) = \partial \bar J^{ij}(\bar z) = 0$. 
Calculating the $JJ$ \ac{OPE} of these currents shows that the system of $2N$ free fermions provides a level $k=1$ realisation of a Ka{\v c}-Moody current algebra\footnote{An excellent review of affine current algebras can be found in \cite{Goddard:1986bp}.} ${\rm SO}(2N)_L\times {\rm SO}(2N)_R$.

We now focus on the left-movers, since the same analysis can be straightforwardly performed on the right-moving CFT as well.
It is convenient to complexify the real fermions into pairs
\begin{align}
	\Psi^{a,\pm}(z) = \frac{1}{\sqrt{2}}(\psi^{2a}(z) \pm \ii \psi^{2a-1}(z)) \,,
\end{align}
where $a=1,\ldots,N$ runs over the rank of the SO$(2N)$ symmetry, and similarly for the right-movers. It is possible to bosonize the left-moving \ac{CFT} of the $N$ complexified fermions $\Psi^{a,\pm}(z)$, that is, to consistently replace it with a system of compact scalar fields $H^a(z)$. Normalizing their 2-point correlators on the sphere as
\begin{align}
	\langle H^a(z)\,H^b(w)\rangle = -\frac{\alpha'}{2}\,\delta^{ab} \log(z-w)\,,
\end{align}
and, adopting for the time being the \ac{CFT} convention $\alpha'=2$, the chiral bosonization corresponds to the identification\footnote{Although the precise relation between $\Psi$ and $H$ is complicated and non-local \cite{Mandelstam:1975hb}, a simple method of bosonization \cite{Friedan:1985ge} can be obtained, by introducing appropriate cocycle phases that ensure the correct fermion anti-commutation relations. While we suppress cocycles in this presentation, an excellent discussion of the technique, particularly useful for concrete calculations, can be found in \cite{Kostelecky:1986xg}.}
\begin{align}
	\Psi^{a,\pm}(z) = :\ee^{\pm iH^a(z)}: \quad,\quad J^a(z) =\ii\partial H^a(z) = :\Psi^{a,+}\Psi^{a,-}:(z) \,,
	\label{FFF:bosoniz}
\end{align}
and we henceforth suppress the explicit display of the normal ordering symbols. It is straightforward to check that the $J^a(z)$ currents defined in  \eqref{FFF:bosoniz} correspond to the $N$ Cartan currents of the Ka{\v c}-Moody algebra SO$(2N)$, while the remaining $4\binom{N}{2}$ currents can be obtained as the combinations $\ee^{\pm \ii H^a(z) \pm \ii H^b(z)}$. Furthermore, it can be checked that the bosonization \eqref{FFF:bosoniz} is consistent with the conformal weights, reproduces the same \ac{OPE}s as the original fermion system, and is therefore equivalent to the \ac{CFT} of the $N$ free compact scalars $H$ with the same left-moving central charge $c_L=N$. 

The (radial) quantization of the free fermion system, of course, depends on the boundary conditions along the non-trivial cycle of the cylinder, which can be Neveu-Schwarz (anti-periodic) or Ramond (periodic). In particular, the Ramond vacuum is degenerate and transforms as an SO$(2N)$ spinor, whose irreducible Weyl representations of opposite chirality will be denoted as $S$ (spinor) and $C$ (conjugate spinor). The associated spin fields $S(z)$ and $C(z)$ are then constructed such that they generate the Ramond vaccum, upon acting on the SL$(2;\mathds C)$ invariant vacuum.

These spin fields admit a simple free-field representation in terms of the bosonization fields $H$. Indeed, both the $\psi(z) S(0)$ and $\psi(z)C(0)$ \ac{OPE}s involve branch cuts in $z$ which identify the helicity charges $q_a, q_a'$ of the spin fields as $q_a, q_a' \in \{\pm\frac{1}{2}\}$,
\begin{align}
	S(z) = \ee^{\ii q_a H^a(z)} \quad,\quad C(z) = \ee^{\ii q'_a H^a(z)} \,.
\end{align}
In this helicity basis, the representation is encoded into the choice of weight vectors $q_a$ and $q'_a$, and the Weyl condition requires keeping an even (odd) number of minus signs in the helicity charges $q_a$, and an odd (even) one in $q_a'$. Note, furthermore, that the conformal weights of $S(z)$ and $C(z)$ are obtained as the squared lengths of the weight vectors, and yield $q\cdot q/2 = q'\cdot q'/2 = N/8$. In particular, for $N=1$ this implies that the Ramond vacuum of a single complex fermion carries conformal weight $\frac{1}{8}$, in accordance with the fact that the $c=\frac{1}{2}$ Virasoro algebra contains two irreducible representations of highest conformal weight $\frac{1}{16}$, corresponding to the ground state of a single real fermion with periodic boundary conditions.

The mode expansion of a free non-compact scalar $\Phi(z,\bar z)$ carries continuous momentum modes $p_L=p_R$ which are always left-right matched, hence, making it incompatible with the holomorphic factorization of the free fermion \ac{CFT}. It is then clear that the chiral bosons $H^a(z)$ must be necessarily compact in order to consistently bosonize the free fermion system. Consider for simplicity a single chiral boson $H(z)$, compactified on a circle of radius $R$. Reinstating the $\alpha'$-dependence in \eqref{FFF:bosoniz}
\begin{align}
	\Psi^{\pm}(z) = \ee^{\pm \ii \sqrt{\frac{2}{\alpha'}}\,H(z)} \quad,\quad J(z) = \ii\sqrt{\frac{2}{\alpha'}}\, \partial H(z) = :\Psi^{+}\Psi^{-}:(z) \,,
	\label{FFF:bosoniz2}
\end{align}
and, requiring that the bosonized representation for $\Psi^{\pm}(z)$ be single valued under $H\to H+2\pi R$, we obtain the $R=\sqrt{\alpha'/2}$. This implies that the fermionization of string coordinates is only consistent at specific points in moduli space, known as the \emph{fermionic point}.

We are now ready to describe the bosonization procedure at the level of the string partition function. At genus one, we must specify the boundary conditions of fields along both non-trivial cycles of the worldsheet torus $z\sim z+1 \sim z+\tau$ of complex structure $\tau=\tau_1+\ii\tau_2$. Consider the chiral complexified fermion $\Psi^{\pm}(z)$ with boundary conditions
\begin{align}
	\begin{split}
		&\Psi^{\pm}(z+1) = -\ee^{\mp \ii\pi \gamma}\,\Psi^{\pm}(z) \,, \\
		&\Psi^{\pm}(z+\tau) = -\ee^{\pm \ii\pi \delta}\,\Psi^{\pm}(z) \,,
	\end{split}
\end{align}
where $\gamma,\delta$ are real parameters twisting the boundary conditions. Note that the special values $\gamma=0,1$ correspond to anti-periodic and periodic boundary conditions along the $a$-cycle of the torus, respectively. Similarly, $\delta=0,1$ assign anti-periodic and periodic boundary conditions along the $b$-cycle. Of course, for a single real fermion $\psi(z)$, the assignments $\gamma,\delta=0,1$ would be the only distinct boundary conditions allowed by the $\mathds Z_2$ automorphism $\psi\to-\psi$ of the left-moving \ac{CFT}. In the case of a complexified fermion $\Psi^{\pm}(z)$, the \ac{CFT} now enjoys a continuous U(1) symmetry, which allows for arbitrary twistings $\gamma,\delta$ of the boundary conditions.

The fermionic path integral essentially amounts to the evaluation of the determinant of a chiral Dirac operator $\bar\partial$ twisted by $\gamma,\delta$. Up to an irrelevant phase, this reads
\begin{align}
	\begin{split}
	{\rm Det}_{\gamma,\delta}(\bar\partial) &=  \ee^{\ii\pi\frac{\gamma\delta}{2}} q^{\frac{\gamma^2}{8}-\frac{1}{24}}\prod_{n>0}\left(1+q^{n+\frac{\gamma}{2}-\frac{1}{2}}\ee^{\ii\pi\delta}\right)\left(1-q^{n-\frac{\gamma}{2}-\frac{1}{2}}\ee^{-\ii\pi\delta}\right) \\
			&= \frac{\vartheta\left[\gamma\atop\delta\right](0;\tau)}{\eta(\tau)}  \,,
	\end{split}
\end{align}
and matches the expression for the partition function of a complex chiral fermion obtained by canonical quantization\footnote{Our conventions are a modified version of \cite{Alvarez-Gaume:1986rcs}. See the same reference for a detailed discussion of the properties of chiral Dirac determinants on Riemann surfaces.}. Here, $\eta$ is the Dedekind function
\begin{align}
	\eta(\tau) = q^{\frac{1}{24}}\prod_{n>0}(1-q^n)\,,
\end{align} 
expressed as a function of the nome $q\equiv \ee^{2\pi \ii\tau}$, while $\vartheta\left[\gamma\atop \delta\right](z;\tau)$ are the Jacobi theta functions with characteristics, with sum representation
\begin{align}
	\vartheta\left[\gamma \atop \delta\right](z;\tau) = \sum_{n\in \mathds Z} q^{\frac{1}{2}\left(n-\frac{\gamma}{2}\right)^2} \,\ee^{2\pi \ii \left(z-\frac{\delta}{2}\right)\left(n-\frac{\gamma}{2}\right)} \,.
	\label{FFF:thetaF}
\end{align}
For simplicity, theta constants $\vartheta\left[\gamma \atop \delta\right](0;\tau)$ will be henceforth denoted simply as $\vartheta\left[\gamma \atop \delta\right]$.

Now consider the \ac{CFT} of a single complex fermion $\Psi^{\pm}(z)$ with boundary conditions $\left[\gamma \atop \delta\right]$. Together with its right-moving counterpart $\bar\Psi^{\pm}(\bar z)$ and, assuming the same boundary conditions, the contribution to the partition function reads
\begin{align}
	Z_{\Psi\bar\Psi}\left[\gamma \atop \delta\right]=\frac{\vartheta\left[\gamma \atop \delta\right] \, \bar\vartheta\left[\gamma \atop \delta\right]}{\eta\bar\eta} 
			=\frac{1}{\eta\bar\eta}\,\sum_{m,n\in\mathds Z} q^{\frac{1}{2}\left(m-\frac{\gamma}{2}\right)^2}\,\bar q^{\frac{1}{2}\left(n-\frac{\gamma}{2}\right)^2} \ee^{-\ii\pi\delta\left(m-n\right)} 
	\label{FFF:lattBoz1}
\end{align}
where we explicitly made use of the sum representation \eqref{FFF:thetaF} of theta constants. Shifting the summation variable $n\to m-n$, we can rearrange the exponents as
\begin{align}
	Z_{\Psi\bar\Psi}\left[\gamma \atop \delta\right]
			=\frac{1}{\eta\bar\eta}\,\sum_{m,n\in\mathds Z} \ee^{-\ii\pi n \delta} \, q^{\frac{\alpha'}{4}\left(\frac{m-n/2-\gamma/2}{\sqrt{\alpha'/2}}+\frac{n}{\sqrt{2\alpha'}}\right)^2}\,\bar q^{\frac{\alpha'}{4}\left(\frac{m-n/2-\gamma/2}{\sqrt{\alpha'/2}}-\frac{n}{\sqrt{2\alpha'}}\right)^2} \,,
	\label{FFF:lattBoz2}
\end{align}
and we recognize on the r.h.s. the shifted (1,1) lattice
\begin{align}
	\hat \Gamma_{1,1}\left[\gamma \atop \delta\right](R) = \sum_{m,n\in\mathds Z} \ee^{-\ii\pi n \delta} \, q^{\frac{\alpha'}{4}\left(\frac{m-n/2-\gamma/2}{R}+\frac{nR}{\alpha'}\right)^2}\,\bar q^{\frac{\alpha'}{4}\left(\frac{m-n/2-\gamma/2}{R}-\frac{nR}{\alpha'}\right)^2} \,,
\end{align}
at the fermionic radius $R=\sqrt{\alpha'/2}$. Summing over all spin structures $\left[\gamma \atop \delta\right]$ in \eqref{FFF:lattBoz2}, we recover the partition function of a scalar compactified on a circle $S^1$ of the same radius. Indeed, the sum over $\gamma$ effectively resets $2m-\gamma \to m$ to integer values, while the sum over $\delta$ projects onto even windings $n\to 2n$,
\begin{align}
	\tfrac{1}{2}\sum_{\gamma,\delta\in\mathds Z_2} Z_{\Psi\bar\Psi}\left[\gamma \atop \delta\right]
			=\frac{1}{\eta\bar\eta}\,\sum_{m,n\in\mathds Z} q^{\frac{\alpha'}{4}\left(\frac{m}{\sqrt{\alpha'/2}}+\frac{n}{\sqrt{2\alpha'}}\right)^2}\,\bar q^{\frac{\alpha'}{4}\left(\frac{m}{\sqrt{\alpha'/2}}-\frac{n}{\sqrt{2\alpha'}}\right)^2} = \frac{\Gamma_{1,1}\left(\sqrt{\tfrac{\alpha'}{2}}\right)}{\eta\bar\eta} \,,
	\label{FFF:lattBoz3}
\end{align}
where, again, on the r.h.s. we recognize the (unshifted) Narain lattice \cite{Narain:1985jj,Narain:1986am}
\begin{align}
	 \Gamma_{1,1}(R) = \sum_{m,n\in\mathds Z}  q^{\frac{\alpha'}{4}P_L^2}\,\bar q^{\frac{\alpha'}{4}P_R^2} \,,
\end{align}
at the fermionic radius, expressed in terms of the left and right moving compact momenta $P_{L,R}= \frac{m}{R}\pm\frac{nR}{\alpha'}$, with $m,n\in \mathds Z$ denoting the Kaluza-Klein momentum and winding numbers, respectively. Analogous expressions can be obtained in the case of several complexified fermions, giving rise to higher dimensional lattices. We will return to this point when we discuss the connection of the \ac{FFF} to orbifold theories in Section \ref{sec:FFForbifold}.


\section{Supersymmetry among free fermions}
\label{sec:FFFSUSYModular}

The equivalence between the worldsheet \ac{CFT}s of free fermions and free scalars compactified at the fermionic radius $R=\sqrt{\alpha'/2}$ opens up the possibility of constructing consistent string theories in which all worldsheet (internal) degrees of freedom are consistently fermionized at special points in moduli space. Embedding this procedure in the string worldsheet is, however, far from trivial. On the one hand, in the case of the superstring, the fermionic degrees of freedom we introduce must realize a local $\mathcal N=1$ superconformal algebra, necessary for the consistent coupling of the theory to two-dimensional gravity and for projecting out unphysical states. On the other hand, not all choices of boundary conditions for the free fermions are consistent with modular invariance at one and higher genera. In this section, we will begin discussing these requirements and the constraints they imply\footnote{For more details, see \cite{Antoniadis:1985az}.}.

Consider a theory of $N$ free left-moving Majorana-Weyl fermions $\psi^i$ with action
\begin{align}
	S = \frac{1}{2\pi} \int \dd^2 z~\psi^i \bar\partial \psi^i \,,
\end{align}
enjoying a global O$(N)$ symmetry, with $i=1,2,\ldots, N$. Observe that the action is invariant under the non-linear supersymmetry transformation
\begin{align}
	\delta\psi^i = \epsilon \,C^{ijk} \psi^j \psi^k \,,
\end{align}
if and only if $C^{ijk}$ is fully antisymmetric \cite{Goddard:1984hg,DiVecchia:1984nyg}. Constructing the generator of this transformation leads to the fermionic realization of the worldsheet supercurrent
\begin{align}
	T_F(z) = \tfrac{1}{3} \ii C^{ijk}\psi^i\psi^j\psi^k(z) \,,
	\label{FFF:supercurrent1}
\end{align}
which indeed carries the correct $(0,\frac{3}{2})$ conformal weight. We should now impose that the $T_F(z)$ in \eqref{FFF:supercurrent1} indeed closes an $\mathcal N=1$ superconformal algebra, i.e. that it satisfies the \ac{OPE}
\begin{equation}
	T_F(z)T_F(w) = \frac{\hat c}{(z-w)^3} + \frac{2}{z-w}\,T(z) + \ldots\,,
	\label{FFF:N=1OPE}
\end{equation}
where $\hat c = 2c/3$, with $c=N/2$ being the central charge of the free fermion \ac{CFT}. Calculating the same \ac{OPE} using the explicit form of the supercurrent \eqref{FFF:supercurrent1}, one obtains
\begin{align}
	\begin{split}
	T_F(z)T_F(w) &= \tfrac{2}{3}\, \frac{C^{ijk}C^{ijk}}{(z-w)^3} + 2C^{ijk}C^{ij\ell}\frac{:\psi^k\psi^\ell :(w)}{(z-w)^2} \,\\
		& -2C^{ijk}C^{ij\ell}\frac{:\psi^k\partial\psi^\ell:(w)}{z-w} - C^{ijk}C^{imn} \frac{:\psi^j\psi^k\psi^m\psi:^n(w)}{z-w} + \ldots
	\end{split}
\end{align}
The requirement that this matches \eqref{FFF:N=1OPE} implies two independent conditions \cite{Antoniadis:1985az}
\begin{align}
\label{FFF:susyCond1}
		&C^{[ij|m}C^{k\ell]m} = 0 \,,\\
		&C^{ijk}C^{ij\ell} = \tfrac{1}{2}\delta^{k\ell} \,. \label{FFF:susyCond2}
\end{align}
The first is a Jacobi identity, implying that $C^{ijk}$ are structure constants of a Lie algebra corresponding to a group $G$, whereas the second implies that $G$ be semi-simple and compact. In other words, requiring the realization of an $\mathcal N=1$ superconformal field theory implies that the $N$ real fermions $\psi^i$ must transform in the adjoint representation of a Lie group, such that the global SO$(N)$ symmetry is gauged into a local symmetry $G$, satisfying  $\dim{G}=N$. We will henceforth write the structure constants in the conventional normalization
\begin{align}
	C^{ijk} = \frac{1}{2\sqrt{h^\vee}} \,f^{ijk}\,,
\end{align}
where $h^{\vee} = f^{ijk}f^{ijk}/2\,{\rm dim}(G)$ is the dual Coxeter number\footnote{Note the relation $c(G)=2h^\vee$, with $c(G)$ being the quadratic Casimir in the adjoint representation of $G$.} of $G$.

Consider now the left-moving sector of superstring theory which, as mentioned above, must enjoy at least $\mathcal N=1$ worldsheet supersymmetry. 
Denoting the number of non-compact dimensions as $D$, and assuming that the internal space is realized entirely in terms of $N$ real free fermions with SO$(N)$ global symmetry, it is clear that the cancellation of the conformal anomaly requires the vanishing of the total central charge
\begin{equation}
	\frac{3}{2}D + N/2 -15 = 0\,,
\end{equation}
with $3D/2$ being the contribution of the $D$ non-compact super-coordinates, $N/2$ that of the system of auxiliary real fermions, and $-15$ being the net contribution of the $b,c,\beta,\gamma$ (super)ghost systems. 

Note that, although it is in many cases possible (and useful) to use bosonization in order to reinterpret the system of $N$ auxiliary fermions in terms of (super)coordinates compactified at the fermionic radius, the idea of the fermionic construction is that it is possible to directly construct consistent string theories in $D<10$ spacetime dimensions, by balancing the central charge deficit created by the lack of a traditional compactification space, against the system of $N=3(10-D)$ worldsheet fermions. Moreover, we shall see that doing so, allows for a complete classification of string models constructible in the free fermionic framework in terms of simple constraints for the boundary conditions of the fermion system and its associated generalized \ac{GSO} projections.

Specializing to $D=4$ dimensions, we see immediately that a total of $N=18$ auxiliary real fermions need to be introduced into the left-moving \ac{CFT}. 
If all 18 real fermions share the same boundary conditions, the unbroken global symmetry group is SO$(18)$ and this has to be gauged down to the compact, semi-simple gauge group $G$. The only such groups of dimension 18 are SU$(2)^6$, SU$(3)\times$SO$(5)$ and SU$(4)\times$SU$(2)$, each realizing a super-Ka{\v c}-Moody current algebra at level $k=h^\vee$ whose $JJ$ \ac{OPE} reads
\begin{align}
	J^i(z)J^j(w) = \frac{k\delta^{ab}}{(z-w)^2} + \ii f^{ijk} \frac{J^k(w)}{z-w}+\ldots 
\end{align}
with the correct central charge 
\begin{align}
	c=\frac{k\,{\rm dim}(G)}{k+h^\vee} = \frac{N}{2}\,.
\end{align}
This closure of the $\mathcal N=1$ super conformal algebra can be explicitly verified by taking the currents $J^i(z) =\frac{1}{2}f^{ijk}\psi^j\psi^k(z)$ as the superpartners of the free fermions $\psi^i$. For instance, in the maximal rank case SU$(2)^6$, we have a $k=2$ realization of the current algebra and the corresponding central charge $c=6\times 3/2$ precisely reflects a system of 3 free fermions for each SU(2) factor.

It is well known \cite{Green:1986dp,Friedan:1984rv,Banks:1987cy}  that in order for the left (resp. right) moving sector of string theory to include massless spacetime fermions, the corresponding left (right) moving super-Ka{\v c}-Moody algebra must necessarily be abelian. For instance, in Type II theories with unbroken spacetime supersymmetry, both the left and right moving current algebras are abelian and, hence, non-abelian interactions may arise only by introducing D-branes. In Heterotic theories, instead, only the left-moving sector enjoys worldsheet supersymmetry and, thus, it is only the left-moving current algebra that is constrained to be abelian. Indeed, the right-moving worldsheet is bosonic, and its current algebra can give rise to non-abelian currents, which translate to the presence of non-abelian gauge fields in the massless spectrum, while the spin fields making up the vertex operator of massless gravitini arise from the left-moving sector. In what follows we focus the analysis entirely on the Heterotic string.

We will now pick the maximal rank case, with local gauge symmetry SU$(2)^6$ and further gauge it down to its abelian factors U$(1)^6$, by an appropriate choice of boundary conditions on the free fermions. Such gaugings down to a subgroup $H$ are consistent, provided $G/H$ is a symmetric space \cite{Antoniadis:1985az}. Consider each SU$(2)$ triplet of fermions $\{\chi^I, y^I, \omega^I\}$ with $I=1,\ldots,6$, such that the left-moving worldsheet supercurrent takes the form
\begin{align}
	T_F(z) = \ii \psi^\mu\partial X^\mu(z) + \ii \sum_{I=1}^{6} \chi^I y^I \omega^I (z) \,,
	\label{FFF:supercurrent2}
\end{align}
where the first term is due to the non-compact super-coordinates carrying the four-dimensional Lorentz indices, while the second term is due to the system of 18 free fermions. It is often convenient to intuitively think of $y^I$ and $\omega^I$ as the auxiliary fermions arising from the fermionization of  6 internal bosonic coordinates $\ii\partial X^I = y^I \omega^I$, compactified on circles of radius $R=\sqrt{\alpha'/2}$. In this sense, $\chi^I$ then plays the role of their fermionic superpartner.

As we have mentioned, breaking the non-abelian local symmetry down to U$(1)$s can be accomplished by assigning different boundary conditions to the fermions $\chi^I, y^I,\omega^I$, realizing each of the SU$(2)$s. However, this assignment is highly constrained by the form \eqref{FFF:supercurrent2} of the full worldsheet supercurrent, which must have well-defined periodicities in order for the $\mathcal N=1$ superconformal theory to remain intact. Indeed, the first term in \eqref{FFF:supercurrent2} is proportional to the worldsheet fermions $\psi^\mu$ transforming under the Lorentz group. If we denote their boundary conditions as $\left[a \atop b\right]$, then the internal part $\chi^I y^I\omega^I$ should also carry the same net (anti)periodicities\footnote{Strictly speaking, the supercurrent $T_F(z)$ in \eqref{FFF:supercurrent2} should also include the contributions of the (super)ghost systems. We do not explicitly display these contributions here, since they will not play any particular role in our analysis. Alternatively, we may work in the lightcone and restrict $\psi^\mu$ and $X^\mu$ to the transverse directions only.}.


\section{Constraints from modular invariance}
\label{sec:FFFSUSYModular}

We have already mentioned that not all spin-structure assignments for the worldsheet fermions are allowed on topologically non-trivial worldsheets. Indeed, the vacuum amplitude of the theory at one and higher genera should remain invariant under large diffeomorphisms. Each such assignment of spin-structures consistent with (multi-loop) modular invariance gives rise to an a priori different string vacuum. These constraints first considered in \cite{Kawai:1986va,Kawai:1986ah} were solved in full generality in \cite{Antoniadis:1986rn} and \cite{Antoniadis:1987wp}, treating both the cases of real and complex fermions, and will be presented in Section \ref{sec:FFFconstructionRules}.

In this section, we discuss the conditions imposed by modular invariance in a simple heterotic setup where all fermions can be complexified. Although this is not the most general case, it is sufficient for outlining the salient features of the construction. The main idea is to impose modular invariance of the vacuum amplitude at one and higher genera, together with factorization, in order to extract conditions on the spin-structure dependent coefficients. 

At genus one, the string worldsheet has the topology of a torus with Teichm\"uller parameter $\tau=\tau_1+\ii \tau_2$ to be integrated over the fundamental domain $\mathcal F=\mathcal H/{\rm SL}(2;\mathds Z)$ with $\mathcal H$ being the upper half-plane. Omitting an overall normalization constant, which is irrelevant for our discussion, the amplitude has the generic form
\begin{align}
	F_1 = \int_{\mathcal F} \frac{\dd^2\tau}{\tau_2}~\sum_{\rm spin\atop structures}N\left[^{\textbf{a}}_{\textbf{b}}\right]\, Z_{b,c}\,Z_{\rm bos}\,Z_{\beta,\gamma}\left[^a_b\right]\,Z_{\rm long}\left[^a_b\right]\,Z\left[^{\textbf{a}}_{\textbf{b}}\right](\tau,\bar\tau) \,.
\end{align}
We now describe the various contributions entering the above expression. $Z_{b,c}(\tau,\bar\tau)=\eta^2\bar\eta^2$ is the spin-structure independent contribution of the $b,c,\bar b,\bar c$ ghosts, while $Z_{\rm bos}(\tau,\bar\tau) = 1/(\sqrt{\tau_2} \,\eta\,\bar\eta)^4$ is the contribution of the non-compact worldsheet coordinates. As expected, the ghost system cancels the oscillator contributions of the longitudinal worldsheet coordinates $X^0, X^1$. Furthermore,
$Z_{\beta,\gamma}(\tau)$ is the contribution of the superghost system which, aside from a possible phase, exactly cancels against that of the longitudinal fermions $\psi^0, \psi^1$
\begin{align}
	Z_{\rm long}(\tau) = \frac{\vartheta\left[^a_b\right]}{\eta} \,,
\end{align}
since $T_F$, $\psi^\mu$ and the superghosts are required to share the same boundary conditions $\left[^a_b\right]$, in order to preserve the unbroken $\mathcal N=1$ worldsheet supersymmetry. The block
\begin{align}
	Z\left[^{\textbf{a}}_{\textbf{b}}\right](\tau,\bar\tau) = \frac{\vartheta\left[^a_b\right](0;\tau)}{\eta(\tau)}\prod_{A} \frac{\vartheta\left[^{a_A}_{b_A}\right](0;\tau)}{\eta(\tau)} \, \prod_{\bar A} \frac{\bar\vartheta\left[^{a_{\bar A}}_{b_{\bar A}}\right](0;\bar\tau)}{\bar\eta(\bar\tau)} \,.
	\label{FFF:ZetaThetas}
\end{align}
contains the contributions of the remaining left and right moving worldsheet fermions. In particular, the first factor carrying spin-structures $\left[^a_b\right]$ is the contribution of the transverse spacetime directions $\psi^3, \psi^4$. Note that we use here the symbol $\left[^{\textbf{a}}_{\textbf{b}}\right]$ with $\textbf{a}=(a,a_1,a_2,\ldots)$ and $\textbf{b}=(b,b_1,b_2,\ldots)$ to collectively denote the spin structure assignment of all left- and right- moving (complexified) fermions, while the indices $A$ and $\bar A$ in eq. \eqref{FFF:ZetaThetas} run over the left and right moving complex fermions, respectively, except $\psi^\mu$. Finally, the coefficients $N\left[^{\textbf{a}}_{\textbf{b}}\right]$ are independent of the complex structure $\tau$, but depend on the spin-structure assignments, and effectively correspond to choices of \ac{GGSO} projections.

We are now ready to extract the constraints of one loop modular invariance on the coefficients. To simplify the analysis and to set the constraints in their standard form in the literature, it is convenient to first switch to a slightly different convention for the Jacobi theta functions
\begin{align}
	\Theta\left[^a_b\right](z;\tau) = \ee^{-\frac{\ii\pi}{2}ab} \, \vartheta\left[^a_b\right](z;\tau)\,,
	\label{FFF:MTheta}
\end{align}
which has the advantage of simpler periodicity properties in the lower argument
\begin{align}
	\Theta\left[^{~~a}_{b+2}\right](z;\tau) = \Theta\left[^a_b\right](z;\tau) \quad,\quad \Theta\left[^{a+2}_{~~b}\right](z;\tau) = \ee^{-\ii\pi b} \, \Theta\left[^a_b\right](z;\tau) \,.
	\label{FFF:MThetas}
\end{align}
The one-loop vacuum amplitude then reads
\begin{align}
	F_1 = \int_{\mathcal F} \frac{\dd^2\tau}{\tau_2^2} \, \frac{1}{\tau_2\eta^2\,\bar\eta^2} \, \frac{1}{|\Xi|}\sum_{\textbf{a}, \textbf{b}} C\left[^{\textbf{a}}_{\textbf{b}}\right]\, \hat Z\left[^{\textbf{a}}_{\textbf{b}}\right] \,,
	\label{FFF:ZetaThetas2}
\end{align}
where we used \eqref{FFF:MTheta} to convert all $\vartheta$'s into $\Theta$'s, and absorbed phase factors into the new constant coefficients 
\begin{equation}
	C\left[^{\textbf{a}}_{\textbf{b}}\right] = \ee^{\frac{\ii\pi}{2}\textbf{a}\cdot\textbf{b}}\,N\left[^{\textbf{a}}_{\textbf{b}}\right]\,,
\end{equation}
where the dot product in the exponent is defined\footnote{For simplicity, we assumed all fermions are complexified so that only integer powers of thetas appear. It is easy to incorporate the case of half-integer powers, by scaling the corresponding elements of the dot product metric by $1/2$.} in the Lorentzian sense, using the signature $(10,22)$ of the fermion charge lattice. Here, $\hat Z$ is simply given by \eqref{FFF:ZetaThetas} with $\vartheta$'s replaced by $\Theta$'s, and we also divide by the order $|\Xi|$ of the finite additive group\footnote{For (anti) periodic boundary conditions, the group $\Xi$ is simply a direct sum of $\mathds Z_2$ factors.} $\Xi$ of boundary condition vectors $\textbf{a}$. Performing the transformation $\tau\to\tau+1$ in \eqref{FFF:ZetaThetas2}, using the modular transformation properties of the Jacobi theta functions, resetting the summation as $\textbf{b} \to \textbf{b}-\textbf{a}+\textbf{1}$, and comparing with the original expression, it is easy to extract the first modular condition
\begin{align}
	C\left[^{\textbf{a}}_{\textbf{b}}\right] = \ee^{\frac{\ii\pi}{4}(\textbf{a}\cdot\textbf{a}+\one\cdot\one)} \, C\left[^{~~~\textbf{a}}_{\textbf{b}-\textbf{a}+\textbf{1}}\right]\,.
	\label{FFF:modCond1}
\end{align}
We also denote by $\one$ the boundary condition vector corresponding to periodic boundary conditions for all fermions. Note that the factor $\ee^{\frac{\ii\pi}{4} \one\cdot\one}= -1$ in the heterotic string and reflects the transformation of the Dedekind functions\footnote{In type II theories, one instead has $\ee^{\frac{\ii\pi}{4}\one\cdot\one}=+1$, since the left and right moving worldsheets are both super-reparametrization invariant and the Dedekind functions arise in pairs, $\eta^{-12}\bar\eta^{-12}$.}. Similarly, requiring the invariance of \eqref{FFF:ZetaThetas2} under the second modular transformation $\tau\to-1/\tau$ yields the second condition
\begin{align}
	C\left[^{\textbf{a}}_{\textbf{b}}\right] = \ee^{\frac{\ii\pi}{2}\textbf{a}\cdot\textbf{b}} \, C\left[^{~\textbf{b}}_{-\textbf{a}}\right]\,.
	\label{FFF:modCond2}
\end{align}

Cluster decomposition requires the factorization of higher genus amplitudes when the donuts making up the Riemann surface are pulled infinitely far apart. This implies that the spin-structure coefficients at genus $g$ must also factorize into products of one-loop coefficients
\begin{align}
	C\left[^{\textbf{a}_{(1)} ,~\textbf{a}_{(2)}, ~\ldots ,~\textbf{a}_{(g)}}_{\textbf{b}_{(1)} ,~\textbf{b}_{(2)} ,~\ldots ,~\textbf{b}_{(g)}} \right] = C\left[^{\textbf{a}_{(1)}}_{\textbf{b}_{(1)}}\right] \, C\left[^{\textbf{a}_{(2)}}_{\textbf{b}_{(2)}}\right] \ldots C\left[^{\textbf{a}_{(g)}}_{\textbf{b}_{(g)}}\right] \,.
\end{align}
As a result, on a higher genus Riemann surface, Dehn twists acting on each separate donut alone leave the amplitude invariant as a consequence of one-loop modular invariance and factorization. In addition, factorization also ensures that Dehn twists mixing the spin-structures of nearby donuts in the chain can be accounted for by imposing invariance of the two-loop vacuum amplitude under the non-trivial twist
\begin{align}
	\Omega \to \Omega' =  \Omega - \begin{pmatrix} 0 & 1 \\ 1 & 0 \end{pmatrix} \,,
	\label{FFF:2loopModular}
\end{align}
where $\Omega$ is the period matrix of the double torus. The generalization of \eqref{FFF:MTheta} to genus two reads
\begin{align}
	\Theta\left[
			\begin{array}{ccc} \alpha_{(1)} & , & \alpha_{(2)} \\ \beta_{(1)} & , & \beta_{(2)} \\ \end{array}\right](z;\Omega) = \sum_{\textbf{n}\in\mathds Z^2} \ee^{\ii\pi (n-\frac{\alpha}{2})^T\Omega(n-\frac{\alpha}{2})+2\pi \ii(z-\frac{\beta}{2})^T(n-\frac{\alpha}{2})-\frac{\ii\pi}{2}\alpha^T\beta} \,,
\end{align}
where $\alpha = (\alpha_{(1)},\alpha_{(2)})^T$, $\beta=(\beta_{(1)}, \beta_{(2)})^T$ are column vectors carrying the boundary conditions ascribed to the transport properties of fermions along the non-contractible cycles of each of the two donuts and, similarly, $z$ is a column vector of Jacobi parameters. It is then straightforward to obtain the transformation of the genus-2 theta constant ($z=0$) under \eqref{FFF:2loopModular}
\begin{align}
	\Theta\left[
			\begin{array}{ccc} \alpha_{(1)} & , & \alpha_{(2)} \\ \beta_{(1)} & , & \beta_{(2)} \\ \end{array}\right](0;\Omega') = \ee^{-\frac{\ii\pi}{2}\alpha_{(1)}\alpha_{(2)}} \,\Theta\left[
		\begin{array}{c c c}
		\alpha_{(1)} & , & \alpha_{(2)} \\
		\beta_{(1)}-\alpha_{(2)} & ,& \beta_{(2)} - \alpha_{(1)} \\
		\end{array}
	\right](0;\Omega) \,.
\end{align}
Inserting this transformation into the genus 2 generalization of the vacuum amplitude \eqref{FFF:ZetaThetas2} and imposing modular invariance, one finds the third and final condition
\begin{align}
	C\left[^\textbf{a}_\textbf{b}\right] \, C\left[^{\textbf{a}'}_{\textbf{b}'}\right] = \delta_a\,\delta_{a'} \,\ee^{-\frac{\ii\pi}{2}\textbf{a}\cdot\textbf{a}'}\, C\left[^{~~\textbf{a}}_{\textbf{b}+\textbf{a}'}\right] \, C\left[^{~~\textbf{a}'}_{\textbf{b}'+\textbf{a}}\right] \,,
	\label{FFF:modCond3}
\end{align}
where the phases $\delta_a = (-1)^{a}$, $\delta_{a'}=(-1)^{a'}$ arise from the modular transformation properties of the 2-loop worldsheet gravitino determinant, which can be uniquely fixed by requiring that the consistent 10d superstrings satisfy this condition\footnote{In type II theories, $\delta_{a}$ should be identified with the spacetime fermion parity of the theory, $(-1)^{F_L+F_R}$, receiving contributions from both the left and the right movers.}. 

The three conditions \eqref{FFF:modCond1}, \eqref{FFF:modCond2} and \eqref{FFF:modCond3} together impose multi-loop modular invariance and can be  solved in general \cite{Antoniadis:1985az, Antoniadis:1987wp}, ensuring the absence of local anomalies, the correct spin-statistics connection and unitarity. The solution corresponding to rational \ac{CFT}s relevant for model building will be outlined in the next section. Before closing the discussion, however, it is useful to recast the 2-loop condition \eqref{FFF:modCond3} to a reduced form that severely constrains the phase dependence on the a-cycle and b-cycle boundary condition assignments. To this end, setting $\textbf{a}'=\textbf{c}'$ and $\textbf{b}'=-\textbf{a}$ in \eqref{FFF:modCond3}, and using \eqref{FFF:modCond2} to bring $\textbf{a}$ back to the upper characteristic, one finds
\begin{equation}
	C\left[^\textbf{~~a}_{\textbf{b}+\textbf{c}}\right]\,C\left[^\textbf{c}_\textbf{0}\right] = \delta_a \delta_c \,C\left[^\textbf{a}_\textbf{b}\right]\,C\left[^\textbf{a}_\textbf{c}\right]\,.
	\label{FFF:modCond3b}
\end{equation}
Clearly, $C\left[^\textbf{c}_\textbf{0}\right] \neq 0$ for all $\textbf{c}$, otherwise all coefficients would trivially vanish. Plugging $\textbf{b}=\textbf{c}=\textbf{0}$ into the above equation and simplifying, yields $C\left[^\textbf{a}_\textbf{0}\right]=\delta_a$, where we have conventionally set the overall normalization to $C\left[^\textbf{0}_\textbf{0}\right]=1$. Substituting this into \eqref{FFF:modCond3b}, we can then finally write the factorization condition as
\begin{equation}
	C\left[^\textbf{~~a}_{\textbf{b}+\textbf{c}}\right] = \delta_a \,C\left[^\textbf{a}_\textbf{b}\right]\,C\left[^\textbf{a}_\textbf{c}\right]\,.
	\label{FFF:modCond3c}
\end{equation}
This form will be particularly useful when we discuss the connection of the \ac{FFF} to toroidal orbifolds in Section \ref{sec:FFForbifold}.


\section{Model construction rules, spectrum and effective superpotential}
\label{sec:FFFconstructionRules}

The  modular invariance constraints discussed in Section  \ref{sec:FFFSUSYModular} can be solved in terms of a  
set of $N$ basis vectors $B=\{\beta_1,\dots, \beta_N\}$
which encode the boundary conditions of the worldsheet fermions
and a set of phases $c{\beta_i\atopwithdelims[] \beta_j}$, $i,j=1,\dots N$
associated with generalized Gliozzi–-Scherk-–Olive (GGSO) projections , 
called spin structure coefficients.

In four space-time dimensions the standard notation of worldsheet fermionic fields used in the \ac{FFF} model building is as follows.
The left moving fermions comprise 20 real  fields. These are  $\psi^\mu$ which stands for
the two  space-time fermions 
in the light-cone gauge, the real fermions $\chi^{1},\dots,\chi^{6}$ which parameterize the six 
fermionic internal coordinates, and the 12 real fermions 
$y^1,\omega^1,\dots, y^6,\omega^6$ that come from the fermionization of the associated internal bosonic coordinates. 
The right moving fields consist of 12 real fermions $\overline{y}^1,\overline{\omega}^1,\dots, \overline{y}^6,\overline{\omega}^6$
ascribed to the fermionization of the internal bosonic coordinates, and 16 complex fermions denoted 
$\bar{\psi}^1,\dots,\bar{\psi}^5$, $\bar{\eta}^1$, $\bar{\eta}^2$, $\bar{\eta}^3$, $\bar{\phi}^1$, $\dots,\bar{\phi}^8$ for reasons 
that will become apparent later.
As explained in Section \ref{sec:FFFSUSYModular}, worldsheet supersymmetry is nonlinearly realized among the left-moving fermions 
$\chi^I, y^I,\omega^I$. Although there are several such realizations \cite{Antoniadis:1987wp},
we focus on the simplest case exploited in heterotic model building
where the supercurrent takes the form \eqref{FFF:supercurrent2}.
In this description, each basis vector consists of a set of phases, e.g.
\begin{align}
\begin{split}
\beta = \left\{
\alpha\left(\psi^\mu\right), \alpha(\chi^{12}),\alpha(\chi^{34}), \alpha(\chi^{56}),\alpha(y^1),\dots, \alpha(y^6),
\alpha(\omega^1),\dots, \alpha(\omega^6);\right.\\
\alpha(\bar{y}^1),\dots, \alpha(\bar{y}^6),
\alpha(\bar{\omega}^1),\dots, \alpha(\bar{\omega}^6)\,,\\
\left.\alpha(\bar{\psi}^1),\dots, \alpha(\bar{\psi}^6),
\alpha(\bar{\eta}^1),\alpha(\bar{\eta}^2),\alpha(\bar{\eta}^3),
\alpha(\bar{\phi}^1),\dots, \alpha(\bar{\phi}^8)
\right\},
\end{split}
\label{FFF:betas}
\end{align}
portraying the parallel transport properties of the worldsheet fermions
\begin{align}
f\to -\ee^{\ii\pi \alpha(f)} f\,,
\label{FFF:ftrans}
\end{align} 
where $\alpha(f)\in(-1,1]$ are in general fractional numbers, with the special cases $\alpha=0,1$
 corresponding to anti-periodic (NS) and
periodic (R) fermions respectively. The semi-colon in \eqref{FFF:betas} separates left and right moving fermions. 

The partition function of the theory can be expressed as a sum over pairs of spin structures
in an abelian
group $\Xi$ spanned by the basis vectors,
$\Xi=\{\xi\vert \xi = m_1 \beta_1 +\dots+m_n\beta_N, m_i=0,\dots,N_i\}\,$,
with $N_i$ the smallest integer for which $N_i \beta_i=0 \mod 2$,
\begin{align}
\begin{split}
Z= \int_{\cal F} \frac{{\ensuremath{\mathrm{d}}}^2\tau}{\tau_2^3\,{\eta}^{12} \overline{\eta}^{24}} 
\frac{1}{2^N}\sum_{\alpha,\beta\in\Xi}\cc{\alpha}{\beta}
&
\prod_{f\in \text{ real left }} 
{\Theta^{\frac{1}{2}}\left[\alpha(f)\atop\beta(f)\right]}
\prod_{f\in \text{ complex left}} 
{\Theta\left[\alpha(f)\atop\beta(f)\right]}\,\\
&
\!\!\!\!\!\!\!\!\prod_{f\in \text{ real right}} 
{\overline{\Theta}^{\frac{1}{2}}\left[\alpha(f)\atop\beta(f)\right]}
\prod_{f\in \text{ complex right}} 
{\overline{\Theta}\left[\alpha(f)\atop\beta(f)\right]}\,.
\end{split}
\label{FFF:fpart}
\end{align}
In the last expression the products extend over real/complex left/right fermions respectively. Here 
$\Theta\left[\alpha(f)\atop\beta(f)\right]$ is the Jacobi theta function with characteristics, $\eta$
stands for the Dedekind eta function and $\mathcal{F}$ is the fundamental domain.

Basis vectors and spin structure coefficients are subject to constraints imposed by 
modular invariance and factorization of the string amplitudes
\eqref{FFF:modCond1},\eqref{FFF:modCond2},\eqref{FFF:modCond3}.
The basis vectors must satisfy
\begin{align}
\begin{split}
N_{ij}  \beta_i{\cdot}\beta_j = &0 \mod 4\,,\ \text{with}\ N_{ij}=\lcm(N_i,N_j)\,,\\
N_i \beta_i{\cdot}\beta_i = &0 \mod 8\,, \ \text{if $N_i$ is even}\ \,,
\end{split}
\end{align}
along with the requirement that the number of real fermions which are periodic in any combination of four basis vectors 
has to be even. Moreover,
to guarantee well defined periodicity properties of the supercurrent \eqref{FFF:supercurrent2} we have to impose
\begin{align}
\beta_i(\chi^I)+\beta_i(y^I)+\beta_i(\omega^I) = \beta_i(\psi^\mu)\! \mod 2\  \forall\ I=1,\dots,6\,.
\end{align}
The spin structure coefficients $\cc{\beta_i}{\beta_j}$ can be expressed in terms of $N(N-1)/2+1$ independent phases, namely, 
$\cc{\one}{\one}, 
\cc{\beta_i}{\beta_j}, i>j =1,\dots,N$,
where $\one\in\Xi\,$ is the vector with all fermions periodic. Furthermore,
composite spin structure coefficients, as the ones appearing in \eqref{FFF:fpart},  can be reduced utilizing 
\begin{equation}
\begin{split}
\cc{\alpha}{\beta+\gamma}=\delta_\alpha \cc{\alpha}{\beta} \cc{\alpha}{\gamma} \ ,\\
\cc{\alpha}{\beta}=\ee^{\ii \pi (\alpha{\cdot}\beta)/2}\cc{\beta}{\alpha}^\ast\ ,\\
\cc{\alpha}{\alpha}=\ee^{\ii\pi(\alpha{\cdot}\alpha+\one{\cdot}\one)/4} \cc{\alpha}{\one}
\,,
\end{split}
\end{equation}
where $\delta_\alpha= (-1)^{\alpha(\psi^\mu)}$, and the lorenzian dot product is defined as follows
\begin{align}
\alpha\cdot\beta=\left\{\frac{1}{2}\sum_{f \in \{{\text{real}\atop \text{left}}\}}+\sum_{f \in \{{\text{complex}\atop \text{left}}\}}
-\frac{1}{2}\sum_{f \in \{{\text{real}\atop \text{right}}\}}-\sum_{f \in \{{\text{complex}\atop \text{right}}\}}
\right\}a(f)\beta(f)\,.
\end{align}
The Hilbert space of states contributing to \eqref{FFF:fpart} can be recast in the form
\begin{align}
{\cal H} = \bigoplus_{\alpha\in\Xi}\prod_{i=1}^N\left\{\ee^{\ii\pi \beta_i F_\alpha}=\delta_\alpha 
\cc{\alpha}{\beta_i}^\ast\right\}{\cal H}_\alpha\,,
\end{align}
where  ${\cal H}_\alpha$ is the Hilbert space of the sector $\alpha$, and the 
expression in curly brackets 
represents a \ac{GSO} projection that projects onto states satisfying $\mathrm{e}^{i\pi \beta_i 
F_\alpha}=\delta_\alpha  \cc{\alpha}{\beta_i}^\ast$.
Here, 
\begin{align}
\beta_i F_\alpha = \left\{
\sum_{f \in \alpha_L} -\sum_{f \in \alpha_R}
\right\} \beta_i(f) F_\alpha(f)
\end{align}
where $\alpha=\{\alpha_\text{L};\alpha_\text{R}\}$ and $F_\alpha(f)$ stands for the fermion number operator which takes
values $+1$ or $-1$ when acting on $f$ or $f^\ast$, respectively.
When the vacuum is degenerate,  our  convention implies
$F(f)=0$ and $F(f)=-1$  for a state annihilated by $f_0$ and $f_0^\ast$ respectively.

The mass formula for string states in a sector $\alpha=\{\alpha_\text{L};\alpha_\text{R}\}$ is
\begin{align}
M_\alpha^2 &= -\frac{1}{2} + \frac{1}{8}\,\alpha_\text{L}{\cdot}\alpha_\text{L} + N_\text{L}\nonumber
= -1 + \frac{1}{8}\,\alpha_\text{R}{\cdot}\alpha_\text{R} + N_\text{R}\,,
\end{align}
where  $N_\text{L}, N_\text{R}$ stand for (sums of) left/right oscillator frequencies, respectively.  For a fermion 
transforming as in \Cref{FFF:ftrans} the oscillator frequencies
are $[\left(1+\alpha(f)\right)/2 + \text{integer}]$ for $f$ and  
$[\left(1-\alpha(f)\right)/2+\text{integer}]$  for 
$f^\ast$. Using the identity $\delta_0 \cc{0}{b_i}^\ast=\delta_{b_i}$, it can be easily shown that the 
massless 
spectrum always includes a state of the form 
$\psi^\mu_\frac{1}{2} \left(\overline{\partial X}\right)^\mu_1\left|0\right>$ which arises from the $0$-sector (NS)  
and contains the 
graviton, the dilaton and the two-index antisymmetric tensor. Similarly, we can infer that the $0$-sector spectrum
does not depend on the GGSO coefficients, but only on the choice of the basis vectors.
Moreover, it turns out that the presence of space-time \ac{SUSY}, and the absence of tachyons, is ensured by 
including the vector $S=\left\{\psi^\mu, \chi^1,\dots, \chi^6\right\}$ in the defining basis set and choosing the relevant 
spin  structure coefficients such that the associated  gravitino multiplet survives.

The low-energy effective theory of a generic heterotic string model in the \ac{FFF} is a $\mathcal{N}=1$ no-scale 
supergravity \cite{Cremmer:1983bf,Ellis:1983ei,Ellis:1984bm,Lahanas:1986uc}. The K\"ahler potential 
of chiral multiplets is exactly calculable at string {tree}-level at every order in $\alpha'$
\cite{Antoniadis:1987zk,Ferrara:1987tp,Ferrara:1987tp,Ferrara:1987jr,Lopez:1994ej}. Furthermore,
the superpotential of the effective theory is also  calculable order by order in the 
$\alpha'$-expansion \cite{Kalara:1990fb,Kalara:1990sq} at string tree-level and receives no string loop corrections
\cite{Witten:1985bz,Dine:1986vd}. 
Superpotential calculation amounts to evaluating correlation functions of the associated primary fields 
which in the case of unpaired real fermions incorporate Ising fields \cite{DiFrancesco:1987ez} that result in 
nontrivial vanishing of superpotential couplings normally allowed by gauge symmetries. 
For example, trilinear superpotential couplings reduce to correlators involving two or three Ising fields of 
which only the following are nonvanishing
\begin{align}
\begin{split}
\langle{\sigma_\pm \sigma_\pm}\rangle=\langle{f f}\rangle=\langle{\bar{f}\,\bar{f}}\rangle=1\,,\\
\langle{\sigma_+ \sigma_- f}\rangle = \langle{\sigma_+ \sigma_- \bar{f}}\rangle=
1/{\sqrt{2}}\,,
\end{split}
\end{align}
where $\sigma_+,\sigma_-$ and $f/\bar{f}$ refer to the order,
disorder and fermion operators, respectively.
The contribution of the complex fields that pertain to the internal fermionic coordinates $\chi^1,\dots,\chi^6$
associated with a conserved $\text{U}(1)$ current of the $\mathcal{N}=2$ worldsheet supersymmetry algebra, leads  
to 
additional restrictive selection rules for the superpotential couplings \cite{Rizos:1991bm,Lopez:1990wt}.

Let us close this session by presenting an illustrative model. We consider the following ten-element basis
$B=\{\beta_1,\dots,\beta_{10}\}$
\footnote{Here we denote the 16 complex right moving fermions as follows: 
$\overline{\psi}^{1\dots5},\overline{\psi}^{6\dots10},
\overline{\psi}^{11\dots15},\overline{\eta}$.}
\begin{align}
	\begin{split}
		\beta_1=\mathds{1}&=\{\psi^\mu,\
		\chi^{1\dots6},y^{1\dots6},\omega^{1\dots6};\overline{y}^{1\dots6},
		\overline{\omega}^{1\dots6},\overline{\psi}^{1\dots5},\overline{\psi}^{6\dots10},
\overline{\psi}^{11\dots15},\overline{\eta}\}\,,\\
		\beta_2=S&=\{\psi^\mu,\chi^{1\dots6}\}\,,\\
		\beta_{2+i}=e_i&=\{y^{i},\omega^{i};\overline{y}^{i},\overline{\omega}^{i}\}\;,i=1\dots6\,,\\	
		\beta_9=b_1&=\{x^{34},\chi^{56},y^{3456};\overline{y}^{3456},
		\overline{\psi}^{1\dots5},\overline{\eta}\}\,,\\
		\beta_{10}=b_2&=\{\chi^{12},x^{56},y^{1256};\overline{y}^{1256},
		\overline{\psi}^{6\dots10},\overline{\eta}\}\,.
	\end{split}
	\label{FFF:model}
\end{align}	
This defines an $\mathcal{N}=1$ supersymmetric\footnote{Provided we choose the GGSO coefficients 
$\cc{S}{e_i}=-1,i=1,\dots,6$.} ${\text{SO}(10)}^3\times\text{U}(1)$ model, where the three $\text{SO}(10)$ factors are  
associated to 
$\overline{\psi}^{1\dots5}$, $\overline{\psi}^{6\dots10}$ and $\overline{\psi}^{11\dots15}$, respectively, and the 
$\text{U}(1)$ factor is related to $\overline{\eta}$. The untwisted sector $(S)+0$
\footnote{Hereafter, we use a compact notation for sectors contributing
to the same field multiplet, for example $(S)+0$ stands for sectors $S$ and $0$.}
matter spectrum is independent of the GGSO coefficient choice
and can be fully derived using the basis \eqref{FFF:model}. As seen from Table \ref{FFF:ims} it includes two 
vectorials from each $\text{SO}(10)$ with opposite $\text{U}(1)$ charges, three bi-vectorials and six total singlets. 
The untwisted sector spectrum depends on the choice of $\cc{\beta_i}{\beta_j}$. However, there is a systematic way
to derive the full spectrum.
Massless states transforming in the spinorial representations of the three 
$\text{SO}(10)$ factors can arise from the sectors ${\cal S}^1_{p^1q^1r^1s^1}=(S+)b_1+p^1 e_3 + q^1 e_4 + r^1 e_5 + s^1 
e_6$,  ${\cal 
S}^2_{p^2q^2r^2s^2}=(S+)b_2+p^2 e_1 +  q^2 e_2 + r^2 e_5 + s^2 e_6$ and ${\cal S}^3_{p^3q^3r^3s^3}=(S+)b_3+p^3 e_1 + 
q^3 e_2 + r^3 e_3 + s^3 e_4$, 
where $p^I,q^I,r^I,s^I=0,1$ for $I=1,2,3$, and $b_3=b_1+b_2+x$ with $x=\one+S+\sum_{i=1}^6 e_i$. Each of the sectors
${\cal S}^I_{pqrs}, I=1,2,3$ can provide one spinorial, so in total we can have 16 spinorials for each $\text{SO}(10)$ 
group.
However, depending on the choice of the GGSO coefficients the number of spinorials can be reduced. This is most easily 
understood by considering GGSO projections of non-overlapping basis vectors. Note that $e_1\cap{\cal 
S}^1_{pqrs}=e_2\cap{{\cal S}^1_{pqrs}}=\text{\O}$, and, similarly, $e_3\cap{{\cal S}^2_{pqrs}}=e_4\cap{{\cal 
S}^2_{pqrs}}=\text{\O}$, $e_5\cap{{\cal S}^3_{pqrs}}=e_6\cap{{\cal S}^3_{pqrs}}=\text{\O}$. Namely, the relative GGSO 
projections in the sectors ${\cal S}^I_{pqrs}$ are
\begin{align}
\cc{{\cal S}^I_{pqrs}}{e_{2I-1}}^\ast=\cc{{\cal S}^I_{pqrs}}{e_{2I}}^\ast=-1\,.
\end{align}
These projections can reduce the number of spinorials by a factor of four, that leads to four spinorials for each 
$\text{SO}(10)$. The spinorial chiralities can be readily obtained using  appropriate GGSO projections. Notice that
$\one+b_2+e_3 + e_4 + r e_5 + s e_6\cap{\cal S}^1_{pqrs}=\{\psi^\mu,\overline{\psi}^{1\dots5}\}$ and analogously
$\one+b_1+e_1 + e_2+r e_5 + s e_6\cap{\cal S}^2_{pqrs}=\{\psi^\mu,\overline{\psi}^{6\dots10}\}$,
$\one+b_1+e_1 + e_2+r e_3 + s e_4\cap{\cal S}^3_{pqrs}=\{\psi^\mu,\overline{\psi}^{11\dots15}\}$. The associated GGSO 
projections yield the chirality ${\cal\chi}^1_{pqrs}$ of the spinorials of the first $\text{SO}(10)$ gauge symmetry
\begin{align}
\begin{split}
{\cal\chi}^1_{pqrs}=-\text{ch}\left(\psi^\mu\right)\cc{{\cal S}^1_{pqrs}}{\one+b_2+e_3 + e_4 + r e_5 + s e_6}^\ast\,,
\end{split}
\end{align}
where $\text{ch}\left(\psi^\mu\right)$ is the space-time chirality,
and similarly for the chiralities ${\cal\chi}^I_{pqrs}, I=2,3$ of the other two $\text{SO}(10)$ factors.
More particularly for the choice
\begin{align}
\label{FFF:ccm}
\cc{\beta_i}{\beta_j}= (-1)^{u_{ij}}\, , u=
\left(
\arraycolsep=4pt
\begin{array}{cccccccccc}
 1 & 1 & 1 & 1 & 1 & 1 & 1 & 1 & 1 & 1 \\
 1 & 1 & 1 & 1 & 1 & 1 & 1 & 1 & 1 & 1 \\
 1 & 1 & 0 & 0 & 1 & 1 & 1 & 1 & 0 & 0 \\
 1 & 1 & 0 & 0 & 1 & 1 & 0 & 0 & 0 & 0 \\
 1 & 1 & 1 & 1 & 0 & 0 & 1 & 0 & 0 & 1 \\
 1 & 1 & 1 & 1 & 0 & 0 & 1 & 1 & 1 & 0 \\
 1 & 1 & 1 & 0 & 1 & 1 & 0 & 0 & 1 & 1 \\
 1 & 1 & 1 & 0 & 0 & 1 & 0 & 0 & 1 & 0 \\
 1 & 0 & 0 & 0 & 0 & 1 & 1 & 1 & 1 & 0 \\
 1 & 0 & 0 & 0 & 1 & 0 & 1 & 0 & 0 & 1 \\
\end{array}
\right)
\end{align}
we obtain the matter spectrum of Table \ref{FFF:ims}.
\begin{table}[!ht]
\centering
\begin{tabular*}{11cm}{@{\extracolsep{\fill}}lcccccc}
\hline
sector&field&\text{SO(10)}&$\text{SO(10)}$&$\text{SO(10)}$&${\text{U}(1)}$&multiplicity\\
\hline
$(S)+0$&$h_1$&$\mathbf{10}$&$\mathbf{1}$&$\mathbf{1}$&$+1$&1\\
&$\overline{h}_1$&$\mathbf{10}$&$\mathbf{1}$&$\mathbf{1}$&$-1$&1\\
&$h_2$&$\mathbf{1}$&$\mathbf{10}$&$\mathbf{1}$&$+1$&1\\
&$\overline{h}_2$&$\mathbf{1}$&$\mathbf{10}$&$\mathbf{1}$&$-1$&1\\
&$h_3$&$\mathbf{1}$&$\mathbf{1}$&$\mathbf{10}$&$+1$&1\\
&$\overline{h}_3$&$\mathbf{1}$&$\mathbf{1}$&$\mathbf{10}$&$-1$&1\\
&$H_{12}$&$\mathbf{10}$&$\mathbf{10}$&$\mathbf{1}$&$0$&1\\
&$H_{13}$&$\mathbf{10}$&$\mathbf{1}$&$\mathbf{10}$&$0$&1\\
&$H_{23}$&$\mathbf{1}$&$\mathbf{10}$&$\mathbf{10}$&$0$&1\\
&$\Phi_i,i=1,\dots,6$&$\mathbf{1}$&$\mathbf{1}$&$\mathbf{1}$&$0$&6\\
\hline
$(S)+b_1$&$S_1$&$\mathbf{16}$&$\mathbf{1}$&$\mathbf{1}$&$+1/2$&1\\
$(S)+b_1+e_5+e_6$&$S_1'$&$\mathbf{16}$&$\mathbf{1}$&$\mathbf{1}$&$-1/2$&1\\
$(S)+b_1+e_3+e_4+e_5+e_6$&$S_1''$&$\mathbf{16}$&$\mathbf{1}$&$\mathbf{1}$&$+1/2$&1\\
$(S)+b_1+e_3+e_4$&$\overline{S}_1$&$\overline{\mathbf{16}}$&$\mathbf{1}$&$\mathbf{1}$&$+1/2$&1\\
\hline
$(S)+b_2+e_5+e_6$&$S_2$&$\mathbf{1}$&$\mathbf{16}$&$\mathbf{1}$&$+1/2$&1\\
$(S)+b_2+e_1+e_6$&$S_2'$&$\mathbf{1}$&$\mathbf{16}$&$\mathbf{1}$&$-1/2$&1\\
$(S)+b_2+e_1+e_2+e_5+e_6$&$S_2''$&$\mathbf{1}$&$\mathbf{16}$&$\mathbf{1}$&$+1/2$&1\\
$(S)+b_2+e_2+e_6$&$\overline{S}_2$&$\mathbf{1}$&$\overline{\mathbf{16}}$&$\mathbf{1}$&$+1/2$&1\\
\hline
$(S)+b_3+e_1+e_2$&$S_3$&${\mathbf{1}}$&$\mathbf{1}$&$\mathbf{16}$&$+1/2$&1\\
$(S)+b_3+e_1$&$S_3'$&${\mathbf{1}}$&$\mathbf{1}$&$\mathbf{16}$&$+1/2$&1\\
$(S)+b_3+e_4$&$S_3''$&${\mathbf{1}}$&$\mathbf{1}$&$\mathbf{16}$&$+1/2$&1\\
$(S)+b_3+e_2+e_4$&$\overline{S}_3$&${\mathbf{1}}$&$\mathbf{1}$&$\overline{\mathbf{16}}$&$-1/2$&1\\
\hline
\end{tabular*}
\caption{\label{FFF:ims}\it Massless matter states and $\text{SO(10)}^3\times{\text{U}(1)}$ quantum numbers of the model
defined in \eqref{FFF:model},\eqref{FFF:ccm}.
}
\end{table}
Note that the $\text{U}_1$ factor is anomalous. 
As inferred from Table \ref{FFF:ims}
\begin{align}
\text{Tr}\left(\text{U}_1\right)=24
\end{align}

The tree-level superpotential is
\begin{align}
\begin{split}
W_3 &= H_{12} H_{13} H_{23} + H_{12} \left(h_1 \overline{h}_2 + h_2 \overline{h}_1\right)
 + H_{13} \left(h_1 \overline{h}_3 + h_3 \overline{h}_1 \right)
+ H_{23} \left(h_2 \overline{h}_3 + h_3 \overline{h}_2 \right)\\
&+\overline{h}_1\left(S_1^2+{S_1^{''}}^2+\overline{S}_1^2\right)
+\overline{h}_2\left({S_3}^2+{S_3^{'}}^2+{S_3^{''}}^2\right)
+\overline{h}_3\left({S_2}^2+{S_2^{''}}^2+\overline{S}_2^2\right)\\
&+ h_1{S_1^{'}}^2+ h_3{S_2^{'}}^2+h_2 \overline{S}_3^2\,.
\end{split}
\end{align}
This is very similar to the orbifold model constructed in \cite{Aldazabal:1994zk} in the context of ``string GUTs" discussed also in \cite{Barbieri:1994jq}.


\section{$\mathcal{N}=1$ supersymmetric models}
\label{sec:FFFSUSYModels}

The \ac{FFF} has yielded several phenomenologically interesting $\mathcal{N}=1$ 
supersymmetric models, with features close to the \ac{MSSM}.
Among the first models built, and most studied,  are the flipped $\text{SU}(5){\times}\text{U}(1)$ model 
\cite{Antoniadis:1987tv,Antoniadis:1988tt,Antoniadis:1989zy}, the  Pati--Salam model
\cite{Antoniadis:1990hb,Leontaris:1999ce} and the Standard-like model \cite{Faraggi:1989ka}.
These are derived from an $\text{SO}(10)$ embedding {realized} using a  common set of seven basis vectors
\begin{align}
\begin{split}
\beta_1 =\zeta & = \left\{\overline{\phi}^{1\dots8}\right\}\,,\\
\beta_2 = S  &= 
\left\{\psi^\mu,\chi^{1\dots6}\right\}\,,\\
\beta_3 =b_1  &= 
\left\{\psi^\mu,\chi^{12},y^{3\dots6};\overline{y}^{3\dots6},\overline{\psi}^{1\dots5},\overline{\eta}^1\right\}\,,\\
\beta_4 =b_2 & = 
\left\{\psi^\mu,\chi^{34},y^{12},\omega^{56};\overline{y}^{12},\overline{\omega}^{56},\overline{\psi}^{1\dots5}\overline{\eta}^2\right\}\,,
\\
\beta_5 =b_3 & = 
\left\{\psi^\mu,\chi^{56},\omega^{1\dots4};\overline{\omega}^{1\dots4},\overline{\psi}^{1\dots5},\overline{\eta}^3\right\}\,,\\
\beta_6 =b_4 & = 
\left\{\psi^\mu,\chi^{12},y^{36},\omega^{45};\overline{y}^{36},\overline{\omega}^{45},\overline{\psi}^{1\dots5},\overline{\eta}^1\right\}\,,\\
\beta_7 =b_5 & = 
\left\{\psi^\mu,\chi^{34},y^{26},\omega^{15};\overline{y}^{26},\overline{\omega}^{15},\overline{\psi}^{1\dots5},\overline{\eta}^2\right\}\,,
\end{split}
\label{FFF:NAHEe}
\end{align}
where included fermions are periodic while all the rest are  antiperiodic.
One or two additional
vectors and an appropriate set of \ac{GGSO} 
phases are then used to break $\text{SO}(10)$ and define each individual model. The first five vectors in 
\Cref{FFF:NAHEe} 
constitute the so called 
NAHE set \cite{Faraggi:1990ac}. The basis vectors $\beta_1=\zeta,\beta_2=S$ and $\one=\zeta+b_1+b_2+b_3$ give rise to 
a $\mathcal{N}=4$ 
supersymmetric model exhibiting $\text{SO}(28){\times}\text{E}_8$ gauge symmetry with the vector $\zeta$ 
associated with the  gauge group enhancement from $\text{SO}(16)$ to $\text{E}_8$.  The vectors $b_1$ and $b_2$ 
correspond to a  $\mathds{Z}_2{\times}\mathds{Z}_2$ orbifold twist that reduces  space-time supersymmetry to 
$\mathcal{N}=1$, gauge symmetry to 
$\text{SO}(10){\times}\text{SO}(6)^3\times\text{E}_8$, and gives rise to chiral fermions. 
Furthermore, the vectors $b_4$ and $b_5$ further 
break the $\text{SO}(6)^3$ group factor to  $\text{SO}(4)^2{\times}\text{U}(1)$.
In the notation used here the ``observable'' $\text{SO}(10)$ gauge symmetry is associated with 
$\overline{\psi}^{1\dots5}$, 
while 
the  ``hidden'' gauge group factor, $\text{E}_8$, pertains to $\overline{\phi}^{1\dots8}$.

The flipped $\text{SU}(5){\times}\text{U}(1)$ model is derived using the basis  $\{\beta_1,\dots, \beta_7,\beta_8\}$, 
where
\begin{align}
\beta_8 =\alpha & = \Bigl\{y^{46},\omega^{46};\overline{y}^{46},\overline{\omega}^{2346},
\alpha({\overline{\psi}^{1\dots5})=\alpha(\overline{\eta}^{123})=\alpha(\overline{\phi}^{1\dots4}})=1/2,\overline{\phi}^{56}\Bigr\}\,.\nonumber
\end{align}
The additional vector $\alpha$ breaks the gauge symmetry down to 
$G=\text{SU}(5)\times{\text{U}(1)'}\times{\text{U}(1)}^4\times{\text{SU}(4)}\times{\text{SO}(10)}$. 
Three fermion generations and a pair of $\text{SU}(5){\times}\text{U}(1)'$ breaking Higgs fields 
in $\boldsymbol{10}_{\nicefrac{1}{2}}+\boldsymbol{\overline{5}}_{\nicefrac{-3}{2}}+\boldsymbol{1}_{\nicefrac{5}{2}}$ 
and $\boldsymbol{10}_{\nicefrac{1}{2}}+\boldsymbol{\overline{10}}_{\nicefrac{-1}{2}}$
representations, respectively,
arise from $b_1,\dots,b_4$ and $b_5$ sectors. Four pairs of $\text{SU}(5)$ {vectors} which accommodate the \ac{MSSM} 
breaking Higgs fields come from the sectors $0$ and $S+b_4+b_5$. The massless spectrum also comprises five hidden sector
multiplets transforming as 
$(\boldsymbol{6},\boldsymbol{1})+(\boldsymbol{1},\boldsymbol{10})$ 
under $\text{SU}(4){\times}\text{SO}(10)$ 
coming from the sectors $b_i+2\alpha(+\zeta),i=1,\dots,5$ and a number gauge singlets originating from $0$ and 
$S+b_4+b_5$ sectors. In addition, six pairs of exotic fractional charge states in 
{$\boldsymbol{4}_{\pm\nicefrac{5}{4}}+\boldsymbol{\overline{4}}_{\mp\nicefrac{5}{4}}$}
representations arise from the sectors $b_1/b_4\pm\alpha(+\zeta), 
b_1+b_3/b_4+b_5\pm\alpha(+\zeta), S/b_1+b_2+b_4\pm\alpha(+\zeta)$.

The {\ac{PS}} model is built upon the basis $\{\beta_1,\dots, \beta_7,\beta_8,\beta_9\}$, where
\begin{align}
\beta_8&=b_6=\left\{y^6,\omega^6;\overline{y}^6,\overline{\omega}^6,\overline{\psi}^{1\dots5},\overline{\eta}^{123},
\overline{\phi}^{1\dots4}\right\}\,,\nonumber\\
\beta_9&=\alpha=\left\{y^{46},\omega^{46};\overline{y}^{46},\overline{\omega}^{2346},\overline{\psi}^{123},\overline{\eta}^{12},
\overline{\phi}^{45}\right\}\,,\nonumber
\end{align}
break the gauge group to 
$G=\text{SU}(4){\times}\text{SU}(2)_\text{L}{\times}\text{SU}(2)_\text{R}\times{\text{U}(1)}^4\times{\text{U}(1)'}\times{\text{SU}(8)}$.
The sectors $b_1,\dots,b_5$ yield three chiral families and one pair of \ac{PS} breaking Higgs multiplets transforming 
as 
$(\boldsymbol{4},\boldsymbol{2},\boldsymbol{1})+(\boldsymbol{\overline{4}},\boldsymbol{1},\boldsymbol{2})$ 
and $(\boldsymbol{4},\boldsymbol{1},\boldsymbol{2})+(\boldsymbol{\overline{4}},\boldsymbol{1},\boldsymbol{2})$,
respectively,
under $\text{SU}(4){\times}\text{SU}(2)_\text{L}{\times}\text{SU}(2)_\text{R}$. Four pairs of {\ac{MSSM}} Higgs 
doublets 
accommodated into  
bi-doublets $(\boldsymbol{1},\boldsymbol{2},\boldsymbol{2})$ 
accompanied with four pairs of triplets into the $(\boldsymbol{6},\boldsymbol{1},\boldsymbol{1})$
 representation of the \ac{PS} group come from the sectors $0, S+b_4+b_5$. 
The sectors $b_1(b_4)+\alpha, b_1+b_2(b_5)+b_4+\alpha,b_2+b_3+b_5+\alpha$ give rise to 10 pairs of exotic fractionally charged 
states transforming as 
$(\boldsymbol{1},\boldsymbol{2},\boldsymbol{1})+(\boldsymbol{1},\boldsymbol{1},\boldsymbol{2})$, while 
$S+b_2+b_4+\alpha$ provides a 
pair of exotic fourplets 
$(\boldsymbol{4},\boldsymbol{1},\boldsymbol{1})+(\boldsymbol{\overline{4}},\boldsymbol{1},\boldsymbol{1})$.
In addition, the massless spectrum includes five pairs of $\text{SU}(8)$ {vectors} 
$\boldsymbol{8}+\boldsymbol{\overline{8}}$ 
from 
$b_i+b_6(+\zeta),i=1,\dots,4, 
b_2+b_3+b_5+b_6(+\zeta)$ and a number of non-{abelian} gauge group singlets from $0$ and $S+b_4+b_5$ sectors.

The Standard-like model is described by the basis  $\{\beta_1,\dots, \beta_7,\beta_8,\beta_9\}$,
with
\begin{align}
\beta_8 
&=\alpha=\left\{\psi^\mu,\chi^{12},y^{36}\omega^{45};\overline{y}^{36},\overline{\omega}^{45},\overline{\psi}^{1\dots5},\overline{\eta}^{123},
\overline{\phi}^{1\dots3},\overline{\eta}^{12},\overline{\phi}^{1\dots4}\right\}\,,\\
\beta_9 &=\beta=\left\{\psi^\mu,\chi^{34},y^{15},\omega^{26};\overline{y}^{1356},\overline{\omega}^{26},
\alpha(\overline{\psi}^{1\dots5})=\alpha(\overline{\eta}^{123})=\alpha(\overline{\phi}^{1567})=1/2,\overline{\phi}^{34}\right\}
\,,\nonumber
\end{align}
together with a specific set of \ac{GGSO} phases \cite{Faraggi:1989ka}.
The resulting gauge group is 
$G=\text{SU}(3){\times}\text{SU}(2){\times}{\text{U}(1)}_\text{C}{\times}{\text{U}(1)}_\text{L}{\times}{\text{SU}(2)}^2{\times}\text{SU}(3){\times}\text{U}(1)^4$.
The sectors $b_1,b_2$ and $b_3$ produce three chiral generations of quarks and leptons, while three pairs of \ac{MSSM} 
Higgs doublets arise from the sector $0$. Moreover, the massless spectrum comprises a number of non-{abelian}
gauge group singlets some of which when acquiring \acp{VEV} could break 
${\text{U}(1)}_\text{C}{\times}{\text{U}(1)}_\text{L}$ down 
to a linear 
combination that is identified with the weak hypercharge. Additional hidden sector 
$\text{SU}(2)^2\times{\text{SU}(3)}{\times}\text{U}(1)^4$ 
states carrying  only {abelian} {\ac{MSSM}} gauge group charges arise from combinations of the vectors 
$\one,b_i,\alpha,\beta$. 
These include a number of fractional charge exotic multiplets. A variation of this model has been discussed in 
\cite{Faraggi:1991jr}. A different class of Standard-like models not based on the NAHE basis set was discussed in 
\cite{Chaudhuri:1995ve}.

Among the main common characteristics of the aforementioned models is the presence of an anomalous {abelian}
gauge symmetry. At first glance, several $\text{U}(1)$'s appear to be anomalous, however,  after 
proper 
redefinitions, only a single linear combination, $\text{U}(1)_A$, turns out to be anomalous, while all other orthogonal 
combinations are anomaly free. The presence of an anomalous $\text{U}(1)_A$ symmetry could destabilise the vacuum and 
lead 
to supersymmetry breaking unless it is cancelled via the Dine--{Seiberg}--Witten mechanism 
\cite{Dine:1987xk,Dine:1987gj}. This involves \acp{VEV} for a set of charged fields $\varphi_i$ that lead to 
$\text{U}(1)_\text{A}$ symmetry breaking and restabilise the vacuum at one loop. These VEVs should comply with the 
requirements of $F$-flatness and $D$-flatness conditions where the later include an anomalous {abelian} symmetry 
related constraint of the form
\begin{align}
D_\text{A} = \sum_i q_\text{A}^i \left|\varphi_i\right|^2+\xi=0\,\ ,\quad  
\text{with} \ \xi = \frac{1}{192\pi^2}\frac{2}{\alpha'}\,\text{Tr}\,{\text{U}(1)}_\text{A}\,.
\label{FFF:ADF}
\end{align}
This introduces {an} extra scale $\xi$ which is typically one or two orders of magnitude below the string scale.

The phenomenological analysis of the models described above entails finding a suitable solution of {$F$-} and 
{$D$-}flatness 
equations for the non-{abelian} singlet fields and, where applicable, the GUT symmetry breaking Higgs fields. Due to 
\eqref{FFF:ADF} flatness solutions lead to field VEVs of order $\xi$ that usually break additional {abelian} gauge 
symmetries and provide masses for extra vector-like matter multiplets.
The contribution of higher order non-{renormalizable} superpotential terms should be also included in the flatness 
analysis when relevant.
Particular attention must be paid on the requirement to keep a pair of \ac{MSSM} Higgs multiplets light, while 
providing 
sufficiently heavy masses to any additional color triplets that mediate nucleon decay. Another source of nucleon decay 
are baryon number violating operators coming from non-{renormalizable} superpotential interactions 
\cite{Ellis:1990vy,Leontaris:1991te,Leontaris:1991sw}.

Flipped $\text{SU}(5)$ model phenomenology has been studied extensively in the literature 
\cite{Antoniadis:1989zy,Lopez:1989fb,Rizos:1990xn,Lopez:1990yk,Leontaris:1990hy,Ellis:1990vy,Lopez:1991ac,Kalara:1991qh,Ellis:1999ce}
including the possibility of allowing hidden sector fields to develop \acp{VEV}
\cite{Antoniadis:1991fc,Antoniadis:1992hs,Antoniadis:2021rfm,Antoniadis:2020txn}. The phenomenological aspects 
of the Pati--Salam model were discussed in \cite{Antoniadis:1990hb,Leontaris:1999ce} and those of the 
Standard-like models were analyzed in 
\cite{Faraggi:1989ka,Faraggi:1991jr,Faraggi:1991be,Faraggi:1992rd,Faraggi:1993rc,Faraggi:1993su,Faraggi:1995bv,Faraggi:1996pa,Cleaver:1999cj,Cleaver:1999mw,Cleaver:2000aa,Cleaver:2001ab,Faraggi:2001ry,Faraggi:2006qa}
and also in \cite{Cleaver:1997jb,Cleaver:1998im,Cleaver:1998gc,Cleaver:1998sm}. In general, one finds flatness 
solutions that lead to hierarchical quark and lepton masses, with the third family being heavier than the other two, 
because third family masses come from trilinear superpotential terms, while the other two arise from  (relatively 
suppressed)  higher order
nonrenormalizable terms. The derivation of neutrino mass matrices is intricate, as neutrinos generally mix with 
all 
singlet fields in the model, however, in certain flatness solutions neutrinos stay sufficiently light thanks to a 
{generalized} see-saw mechanism \cite{Antoniadis:1992hs,Faraggi:1993zh}.

A common feature of the models presented above is the presence of exotic fractionally charged particles in 
their massless {spectra}. Actually, the appearance of color-singlet fractionally charged states is a generic 
property of 
heterotic string {compactifications} based on level $\text{k}=1$ Ka{\v c}--Moody algebra embeddings of the 
non-{abelian} 
group factors of the {\ac{SM}} \cite{Wen:1985qj,Athanasiu:1988uj,Schellekens:1989qb}. Lacking any experimental 
confirmation and in view of strict cosmological constraints, 
one may wish to get rid of these states.
{This can be achieved if all fractionally charged states are confined.} This scenario can be {realized} in the 
case of the flipped $\text{SU}(5)$ 
model where all exotics transform non-trivially under an $\text{SU}(4)$ hidden group factor 
\cite{Ellis:1990iu,Leontaris:1990bw}. Another interesting 
possibility is to project out all fractional charge exotics from the massless spectrum. As shown in 
\cite{Assel:2009xa,Christodoulides:2011zs,Bernard:2012vf}, this is possible in a class of Pati--Salam models (named 
``exophobic'')  built on the symmetric basis discussed in Section \ref{FFF:subsII}.

Free fermionic models based on different gauge groups than those considered above have been discussed in  
 \cite{Schwartz:1989gh,Cleaver:2000ds,Cleaver:2002ps,Faraggi:2006qa}. Moreover, the possibility of building  higher 
 level Ka{\v c}--Moody algebra models using free fermions have been considered in 
 \cite{Lewellen:1989qe,Chaudhuri:1994cd,Dienes:1996yh}.


\section{The symmetric basis and model scans}
\label{FFF:subsII}
An early attempt to classify heterotic string vacua in the \ac{FFF} using computer-assisted search was reported in 
\cite{Finnell:1995ig}. However, a systematic exploration and classification of \ac{FFF} vacua became possible
after adopting a new approach proposed in \cite{Faraggi:2004rq}.
This is build on an 12 element basis 
\begin{align}
	\begin{split}
		\beta_1=\mathds{1}&=\{\psi^\mu,\
		\chi^{1\dots6},y^{1\dots6},\omega^{1\dots6};\overline{y}^{1\dots6},
		\overline{\omega}^{1\dots6},\overline{\eta}^{1,2,3},
		\overline{\psi}^{1\dots5},\overline{\phi}^{1\dots8}\}\,,\\
		\beta_2=S&=\{\psi^\mu,\chi^{1\dots6}\}\,,\\
		\beta_{2+i}=e_i&=\{y^{i},\omega^{i};\overline{y}^{i},\overline{\omega}^{i}\}\;,i=1\dots6\,,\\	
		\beta_9=b_1&=\{x^{34},\chi^{56},y^{3456};\overline{y}^{3456},
		\overline{\psi}^{1\dots5},\overline{\eta}^1\}\,,\\
		\beta_{10}=b_2&=\{\chi^{12},x^{56},y^{1256};\overline{y}^{1256},
		\overline{\psi}^{1\dots5},\overline{\eta}^2\}\,, \\
		\beta_{11}=z_1&=\{\overline{\phi}^{1\dots4}\}\,,\\
		\beta_{12}=z_2&=\{\overline{\phi}^{5\dots8}\}\,,
	\end{split}
	\label{FFF:symbasis}
\end{align}	
which we refer to 
as symmetric basis in the sense that it treats the left/right internal fermionic coordinates, 
$y^{1\dots6}/\overline{y}^{1\dots6}$ and $\omega^{1\dots6}/\overline{\omega}^{1\dots6}$, symmetrically. The choice 
\eqref{FFF:symbasis}
also induces maximal rank reduction (by 6 units) and breaks gauge symmetry down to 
$G=\text{SO}(10)\times{\text{U}(1)}^3\times{\text{SO}(8)}^2$ for generic values of the GGSO. $\text{SO}(10)$ 
{spinors}, 
 $\mathbf{16}(\mathbf{\overline{16}})$, accommodating chiral matter, arise from the twisted sectors  $\mathcal{
 S}^i_{pqrs}=S+b_i+p e_j+q e_k+ r e_\ell+s e_m$ where $p,q,r,s=0,1$ and $(ijk{\ell}m)=(13456),(21256),(31234)$, with 
 $b_3=b_1+b_2+x, x=\one+S+e_1+{\dots}+e_6+z_1+z_2$. Similarly, $\text{SO}(10)$ {vectors}, $\mathbf{10}$ carrying 
 \ac{MSSM} 
 Higgs
 doublets come from the sectors ${\cal V}^i_{pqrs}={\cal S}^i_{pqrs}+x$. 
 Taking into account
 the fact that each of the aforementioned sectors $\mathcal{S}^i_{pqrs}, \mathcal{V}^i_{pqrs}$ 
 can produce a single $\text{SO}(10)$ {spinor/vector} multiplet, one can perform the \ac{GGSO} projections 
 explicitly and derive analytic formulae 
 for the phenomenological characteristics of models, as the number of generations, the {\ac{MSSM}} Higgs multiplets 
 and the 
 number of exotics, in terms of $\cc{\beta_i}{\beta_j}$.
 For example, the net chirality, $n_g$, that is 
 the number of chiral spinorial representations of $\text{SO}(10)$, $n_{\mathbf16}$, minus the number of anti-spinorials, $ n_{\overline{\mathbf{16}}}$,  is given by
\begin{align}
\begin{split}
	n_g=n_{\mathbf16}-n_{\overline{\mathbf{16}}}&=
\frac{1}{2^4}\sum_{p,q,r,s\in\{0,1\}}\sum_{i=1}^3 \cc{\mathcal{S}^i_{pqrs}}{(irs)}^\ast\times\\
&\prod_{i'=2i-1,2i}\left(1-\cc{\mathcal{S}^i_{pqrs}}{e_{i'}}^\ast\right)
\prod_{k'=1,2}\left(1-\cc{\mathcal{S}^i_{pqrs}}{z_{k'}}^\ast\right)\,,
\end{split}
\end{align} 
where  we use the following notation for the chirality factor $\cc{\mathcal{S}^i_{pqrs}}{(irs)}$: $(1rs)=S+b_2+(1-r) 
e_5+(1-s)e_6$, $(2rs)=(S+b_1+(1-r) e_5+(1-s)e_6$ and $(3rs)=(S+b_1+(1-r) e_3+(1-s)e_4$. Similarly, for the vectorial 
representations of $SO(10)$ we have
\begin{align}
\begin{split}
n_{10}=\frac{1}{2^4}\sum_{i=1}^3
\sum_{p,q,r,s\in\{0,1\}}\prod_{i'=2i-1,2i}\left(1-\cc{\mathcal{V}^i_{pqrs}}{e_{i'}}^\ast\right)
\prod_{k'=1,2}\left(1-\cc{\mathcal{V}^i_{pqrs}}{z_{k'}}^\ast\right)\,.
\end{split}
\end{align}
   The introduction of one or two additional
 basis vectors, involving the 16 right complex fermions solely, further breaks gauge symmetry and truncates
 $\text{SO}(10)$ {spinors/vectors} appropriately. In this case, the above equations are 
 modified accordingly and extended .  
  Although intricate, and not invertible in the sense that one cannot solve analytically for the 
  projection coefficients in terms of the model data e.g. $n_g, n_{10}$, these formulae can be readily evaluated using a computer code and utilized to efficiently scan big classes of \ac{FFF} vacua.

The method outlined above was introduced in \cite{Gregori:1999ny} in the context of classification of Type IIA/B string 
models and has been further developed and employed in 
\cite{Faraggi:2004rq} to classify heterotic $\text{SO}(10)$ vacua. It was also used in \cite{Assel:2010wj,Assel:2009xa} 
to classify a huge collection of $10^{11}$ Pati--Salam string models, 
generated by the introduction of a single additional basis vector 
$\beta_{13}=\alpha=\{\overline{\phi}^{45},\overline{\phi}^{12}\}$. 
It was shown that one in a billion Pati--Salam vacua possesses three generations together with the required gauge symmetry 
breaking Higgs fields and is free of massless fractional charged exotics (exophobic models).  Big collections of 
flipped $\text{SU}(5)$, Standard-like and Left-Right symmetric models have been also studied and {categorized} 
in \cite{Faraggi:2014hqa}, \cite{Faraggi:2017cnh} and \cite{Faraggi:2018hqx} respectively. This framework can be 
employed to incorporate additional constraints, related to superpotential couplings, in model scans, as the existence of 
tree-level top quark mass related coupling in the effective superpotential 
that naturally leads to a large mass for the top quark \cite{Rizos:2014uba,Christodoulides:2011zs}. 

The class of models generated by the symmetric basis \eqref{FFF:symbasis} has been shown to exhibit an  
interesting symmetry called spinor-vector duality. This refers to the interchange of matter representations of the 
$\text{SO}(10)$
group factor $\mathbf{16}$ (spinorial) and $\mathbf{10}$ (vectorial) accompanied by the $\mathbf{1}$ (singlet) 
that appear in the decomposition $\mathbf{27}=(\mathbf{16},1/2)+(\mathbf{10},-1)+(\mathbf{1},+2)$ of 
$\text{E}_6\supset{\text{SO}(10)}$ 
\cite{Faraggi:2006pk,Faraggi:2007ms,Catelin-Jullien:2008ssd,Angelantonj:2010zj,Faraggi:2011aw,Faraggi:2021yck}.
This symmetry has been employed in \cite{Faraggi:2014ica} to construct Pati--Salam models with an extra family 
universal $Z'$ symmetry that could survive down to low energies.

The symmetric basis can also be utilised for the efficient implementation of the \ac{FFF} in the exploration of the string landscape using advanced computational methods, 
such as genetic algorithms \cite{Abel:2014xta}, satisfiability modulo theory \cite{Faraggi:2021mws} and, recently, on quantum annealers \cite{Abel:2023rxo}. 


\section{Non-supersymmetric models}
\label{sec:NonSUSY_FFF_Models}

 As has been known since the early days of the first superstring revolution, besides the space-time supersymmetric
$\text{E}_8\times\text{E}_8/\text{SO}(32)$ heterotic string theory
\cite{Gross:1984dd} one can construct consistent 
non-supersymmetric 
theories as the $\text{SO}(16)\times\text{SO}(16)$ heterotic string model \cite{Alvarez-Gaume:1986ghj,Dixon:1986iz}.
However, string phenomenology has mainly focused on $\mathcal{N}=1$ supersymmetric models for two reasons. The first is 
the presence of tachyons in the physical spectrum of a generic non-supersymmetric theory signaling vacuum instability 
and the second is the appearance of large one-loop dilaton tadpoles and their back-reaction on the classical vacuum.
However, the lack of evidence in favor of supersymmetry in recent experiments motivates the exploration of 
non-supersymmetric string vacua. 

It is not difficult to construct four-dimensional non-supersymmetric string models in the FFF framework. It amounts to choosing the 
spin structure coefficient $\cc{S}{\beta}=+1$ for some basis vector $\beta\cap S = \text{\O}$ so as to project out the 
gravitino state arising in the sector $S=\left\{\psi^\mu, \chi^1,\dots, \chi^6\right\}$.
One may then eliminate tachyons from the physical spectrum by an appropriate choice of GGSO coefficients  $\cc{\beta_i}{\beta_j}$.
Several models have been constructed along these lines
\cite{Chamseddine:1988ck,Faraggi:2007tj,Faraggi:2009xy,Faraggi:2020wld,Faraggi:2020wej,Faraggi:2019drl,Faraggi:2022hut}.
Unfortunately, this direct method generally results in explicit supersymmetry breaking at the string scale. 
Another interesting possibility is the spontaneous breaking of supersymmetry via coordinate dependent compactifications 
\cite{Rohm:1983aq,Kounnas:1988ye,Ferrara:1988jx,Kounnas:1989dk}, corresponding to a stringy realization of the 
Scherk--Schwarz mechanism \cite{Scherk:1978ta,Scherk:1979zr}. In its simplest form, this can be achieved by 
 picking an extra dimension $X^5$ compactified on a circle of radius $R$, and imposing non-trivial monodromies $\Phi(X^5+2\pi R)=\ee^{\ii Q} \Phi(X^5)$ around the circle, so that the states $\Phi$ of the theory are periodic only up to the action of a symmetry generator $Q$. This results 
in a shift of the tower of {\ac{KK} masses}
of the charged states. Identifying the symmetry operator $\ee^{\ii Q}$ with the fermion number parity
$(-1)^F$ leads to spontaneous supersymmetry  breaking at a scale $m_{3/2}\sim 1/R$. In general, the scalar potential of the resulting theories is no-longer super-protected, and quantum corrections will typically generate a non-trivial cosmological constant. In the large radius limit, the one-loop effective potential takes the form \cite{Itoyama:1986ei,Antoniadis:1990ew}
\begin{align}
V_\text{eff} = -\frac{\zeta}{R^4} + {\mathcal{O}}\left(\ee^{-\lambda R}\right)\,,
\label{FFF:SNSC}
\end{align}
where $\zeta\sim n_{\rm B}-n_{\rm F}$ is proportional to the massless spectrum degeneracies and $\lambda$ is a positive constant of 
order one. Unfortunately, this value for the cosmological constant deviates from its observed one by several orders of magnitude, 
even if we were to lower the compactification scale down to the TeV range. However, the leading contribution in \eqref{FFF:SNSC} may vanish
 for models where the number $n_{\rm F}$ of the massless fermionic degrees of freedom equals the number $n_{\rm B}$ of 
 bosonic massless states, namely  $n_\text{B}=n_\text{F}$. In this special class of models, termed super no-scale models in \cite{Kounnas:2016gmz}, 
the cosmological constant is exponentially suppressed for sufficiently large values of the compactification radius.
Such non-supersymmetric models have recently attracted attention  
\cite{Abel:2015oxa,Kounnas:2017mad,Abel:2017rch,Itoyama:2020ifw,Itoyama:2021fwc} in the context of string phenomenology. 
Concrete super no-scale models with an even number of generations were constructed and discussed in 
\cite{Abel:2015oxa,Florakis:2016ani,Florakis:2017ecd,Abel:2017vos,Aaronson:2016kjm,Florakis:2021bws}, while three generation models
with exponentially suppressed cosmological constant were also shown to be possible \cite{Florakis:2022avh}, provided additional constraints are satisfied,  
at least as far as $\mathds{Z}_2\times\mathds{Z}_2$ compactifications are concerned.  

The stringy version of the Scherk-Schwarz mechanism can be conveniently realised in terms of freely-acting $\mathds{Z}_2$ orbifolds with action $g=(-1)^F\,\delta$, where $\delta$ is an order-2 translation (shift) along a non-trivial cycle of the compactification space \cite{Ferrara:1987qp}. A large class of semi-realistic \ac{FFF} models may be recast into an orbifold representation and deformed away from the fermionic point by marginal operators \cite{florakis:tel-00607408,Florakis:2016ani,Florakis:2017ecd,Florakis:2021bws}. This effectively reinstates the dependence on the compactification moduli and allows one to study the conditions under which the supersymmetry breaking is spontaneous (Scherk--Schwarz type) or explicit. The procedure for this map is outlined in the following section, and illustrated with an explicit example.


\section{Map to Orbifolds}
\label{sec:FFForbifold}

The fermionic models we have discussed live in special points of moduli space, where the compactification moduli take the fixed values compatible with bosonization. However, in many applications, it is necessary to reinstate the moduli dependence of the theory, for instance, in order to study the dependence of string threshold corrections to the compactification moduli \cite{Florakis:2017ecd}. Provided the corresponding moduli scalars are not projected out of the string spectrum, and provided their vertex operators are exactly marginal, they can be used to marginally deform the theory away from the fermionic point. This can be most easily accomplished if the free-fermionic models are mapped to toroidal orbifolds with appropriate $(\mathds Z_2)^M$ rotations (or even translations) on the internal space (super)coordinates. It should be mentioned that considerable efforts have been made in the literature to bridge the gap between the \ac{FFF}, orbifolds and geometric formulations and obtain a unified treatment, \emph{c.f.} \cite{Chamseddine:1989mz,Donagi:2008xy,florakis:tel-00607408, Athanasopoulos:2016aws,Florakis:2016ani,Florakis:2017ecd,Florakis:2021bws,Faraggi:2022gkt}. The method for the map we present here appeared in its early form in \cite{florakis:tel-00607408}, and further developed in \cite{Florakis:2023aud}.

For simplicity, consider two bosonic coordinates $X^1,X^2$ compactified on a $T^2$, without any orbifold rotation. They can be fermionized in terms of four (auxiliary) real left-moving fermions $y^1,\omega^1,y^2,\omega^2$, as $\sqrt{2/\alpha'}\partial X^i = y^i\omega^i$ and similarly for the right-movers. Generalizing \Cref{FFF:lattBoz3}, we pick identical boundary conditions for the four fermions and sum the contribution to the partition function over all spin structures. After appropriate shifts of the summation variables, one obtains
\begin{equation}
	\frac{1}{2}\sum_{\gamma,\delta=0,1}\frac{\vartheta^2\left[\gamma \atop \delta\right] \, \bar\vartheta^2\left[\gamma \atop \delta\right]}{\eta^2\bar\eta^2} = \frac{1}{\eta^2\bar\eta^2}\,\Gamma_{2,2}(\ii\alpha',\ii) \,,
	\label{FFF:T2fermioniz}
\end{equation}
where $\Gamma_{2,2}(T,U)$ is the partition function of the (2,2) Narain lattice 
\begin{equation}
	\Gamma_{2,2}(T,U) = \sum_{m_i,n_i\in\mathds Z} q^{\frac{\alpha'}{4}|P_L|^2} \, \bar{q}^{\frac{\alpha'}{4}|P_R|^2} \,,
		\label{FFF:Narain22}
\end{equation}
with the complexified lattice momenta defined as
\begin{equation}
	P_L = \frac{m_2-Um_1+T(n_1+Un_2)}{\sqrt{T_2 U_2}} \quad,\ P_R = \frac{m_2-Um_1+\bar T(n_1+Un_2)}{\sqrt{T_2 U_2}}\,.
	\label{FFF:complexMomenta}
\end{equation}
Here, $T$ and $U$ are the Kahler and complex structure moduli of the $T^2$, respectively. Notice that the fermionization of the bosonic $T^2$ coordinates occurs only for a square lattice, as indicated by the purely imaginary values $T=\ii\alpha'$ and $U=\ii$ in \eqref{FFF:T2fermioniz}. Permissible orbifold actions compatible with the fermionization must act crystallographically on this square lattice, so that $\mathds Z_2$ and $\mathds Z_4$ actions can be realized. This naturally generalizes to orbifolds on higher-dimensional tori, as well as to orbifolds involving both twists (rotations) and shifts (translations). 

Let us assume a simple $\mathds Z_2$ twist under which the bosonic coordinates $X^1, X^2$ of $T^2$ are rotated by an angle $\pi$, such that $X^1\to -X^1$ and $X^2\to -X^2$. In terms of the free fermions, this action is reproduced by the twist
\begin{equation}
	y^1\to -y^1 \quad, \ y^2\to -y^2 \quad, \ \omega^1\to \omega^1 \quad, \ \omega^2\to \omega^2 \,,
	\label{FFF:fermTwistBC}
\end{equation}
with an identical action on the right-movers. Furthermore, in order to preserve $\mathcal N=1$ worldsheet supersymmetry, the orbifold action on the bosonic coordinates $X^1, X^2$ must be accompanied by a simultaneous rotation of their real (left-moving) fermion superpartners, i.e. $\psi^1\to -\psi^1$ and $\psi^2\to -\psi^2$. Of course, this $\mathds Z_2$ action on a single $T^2$ does not preserve any spacetime supercharges. The situation can be remedied by extending the $\mathds Z_2$ action to an additional 2-torus, so that one considers $T^4/\mathds Z_2$. 

We postpone the discussion of the full supersymmetry-preserving orbifold action for later and instead focus on the contribution of a single $T^2$ factor to the partition function in the case $T/\alpha'=U=\ii$. This amounts to computing the genus-one path integral of the compact bosonic coordinates spanning $T^2/\mathds Z_2$. Upon conveniently complexifying $Z=X^1+\ii X^2$, and introducing the $\mathds Z_2$ orbifold parameters $h,g\in\{0,1\}$ encoding the twists of the (initially periodic) boundary conditions of $Z$ along the two cycles of the worldsheet torus,  the boundary conditions read
\begin{equation}
	\begin{split}
		& Z(\sigma^1+1,\sigma^2) = \ee^{\ii\pi h}Z(\sigma^1,\sigma^2) \,,\\
		& Z(\sigma^1,\sigma^2+1) = \ee^{\ii\pi g}Z(\sigma^1,\sigma^2) \,.\\
	\end{split}	
	\label{FFF:ZboundaryCond}
\end{equation}
The general mode expansion compatible with these boundary conditions reads
\begin{equation}
	Z(\sigma^1,\sigma^2) = z_0 + Q \sigma^1 + \tilde Q \sigma^2 + \sum_{m,n\in\mathds Z}Z_{n,m}\,\ee^{2\pi \ii[(n+h/2)\sigma^1+(m+g/2)\sigma^2]}\,,
	\label{FFF:modeExpansionZ}
\end{equation}
where $\sigma^1, \sigma^2$ are the 2d coordinates parametrizing the worldsheet torus, $z_0$ is the (center of mass) zero mode, $Q$ and $\tilde Q$ are (complexified) windings corresponding to the classical solution \Bsquare$Z=0$, while $Z_{n,m}$ correspond to the (quantum) oscillator modes.

Clearly, in the sector $h=g=0$ where $Z$ is untwisted under both cycles and the boundary conditions trivialize, one recovers the r.h.s. of \eqref{FFF:T2fermioniz}. Indeed, in this case the path integral over the oscillator modes simply produces Dedekind functions $(\eta^2\bar\eta^2)^{-1}$, the classical BPS solution produces the Narain lattice sum in the Lagrangian representation, and the integral over the zero mode $z_0$ produces the appropriate $T^2$ volume factor. After Poisson resumming over the $\tilde Q$ windings, the volume factor cancels and one recovers the familiar Hamiltonian form of the Narain lattice \eqref{FFF:Narain22}, in terms of the complexified lattice momenta $P_L,P_R$ of eq. \eqref{FFF:complexMomenta}. 

The contribution of the twisted sectors $(h,g)\neq (0,0)$ is more involved. First, substituting the mode expansion \eqref{FFF:modeExpansionZ} into the boundary conditions \eqref{FFF:ZboundaryCond} requires $(1-\ee^{\ii\pi h})Q=(1-\ee^{\ii\pi g})Q=0$ and similarly for $\tilde Q$, implying the absence of windings in the twisted sectors, $Q=\tilde Q=0$. One is left to impose the center of mass conditions
\begin{equation}
	\begin{split}
		&(1-\ee^{\ii\pi h})z_0 = 0 ({\rm mod}\,\Lambda)\,, \\
		&(1-\ee^{\ii\pi g})z_0 = 0 ({\rm mod}\,\Lambda)\,, \\
	\end{split}
\end{equation}
where ${\rm mod}\, \Lambda$ simply denotes the fact that the above conditions should be satisfied modulo $T^2$ lattice vectors. In other words, the center of mass mode $z_0$ is restricted to lie on the simultaneous fixed points under both $h$ and $g$ twists. Let us now return to the path integral. Integrating out the oscillator modes is straightforward and amounts to computing the zeta-regularized determinant ${\rm Det}'_{h,g}$\Bsquare\ of the worldsheet torus Laplacian under the twisted boundary conditions $(h,g)$, yielding $\left|\eta/\vartheta\left[1-h\atop 1-g\right]\right|^2$. Since $Q=\tilde Q=0$, there is no BPS lattice sum present. Moreover, there is no volume factor, but we instead have to perform a discrete sum over the zero mode $z_0$, which spans the simultaneous orbifold fixed points in the $(h,g)$ sector. For $T^2/\mathds Z_2$ with $z_0\sim z_0+1\sim z_0+\ii$, the four fixed points are given by $z_0\in\{0,1/2,\ii/2,(1+\ii)/2\}$. Putting everything together, the contribution to the partition function of the compact scalars $X^1,X^2$ parametrizing $T^2/\mathds Z_2$ in the $(h,g)$ orbifold sector reads
\begin{equation}
	\frac{1}{\eta^2\bar\eta^2}\,\Gamma_{2,2}\left[h\atop g\right](\ii\alpha',\ii) = \frac{1}{\eta^2\bar\eta^2}\left\{ \begin{array}{c l}
			\Gamma_{2,2}(\ii\alpha',\ii) & ,\ (h,g)=(0,0) \\
			\left| \frac{2\eta^3}{\vartheta\left[1-h\atop 1-g\right]}\right|^2 & ,\ (h,g)\neq (0,0)\\
	\end{array} \right. \,.
	\label{FFF:T2Z2orbTwist}
\end{equation}
From the orbifold perspective, the parameter $h=0,1$ labels the orbifold twisted sectors, while summation over $g=0,1$ imposes the $\mathds Z_2$ projection.

It is now straightforward to see that the same result eq.\eqref{FFF:T2Z2orbTwist} can be equivalently obtained from the free fermion system $y^i,\omega^i, \bar y^i, \bar\omega^i$ with the twisted boundary conditions \eqref{FFF:fermTwistBC}. Indeed, assume that before the twists all auxiliary fermions carried the same boundary conditions $(\gamma,\delta)$ as in \eqref{FFF:T2fermioniz}. Introducing the twist implies the replacement
\begin{equation}
	\frac{1}{2}\sum_{\gamma,\delta=0,1}\frac{\vartheta^2\left[\gamma \atop \delta\right] \, \bar\vartheta^2\left[\gamma \atop \delta\right]}{\eta^2\bar\eta^2} \to \frac{1}{2}\sum_{\gamma,\delta=0,1}\frac{\vartheta\left[\gamma \atop \delta\right]\, \vartheta\left[\gamma+h \atop \delta+g\right] \, \bar\vartheta\left[\gamma \atop \delta\right]\, \bar\vartheta\left[\gamma+h \atop \delta+g\right]}{\eta^2\bar\eta^2} \,,
		\label{FFF:TwistReplacement}
\end{equation}
with the untwisted theta functions corresponding to the untwisted fermions $\omega^i$, while the twisted ones are associated to $y^i$, and similarly for the right-movers. Using Jacobi's triple product identity $\vartheta_2\vartheta_3\vartheta_4=2\eta^3$, it is straightforward to check that the free-fermionic partition function on the r.h.s. of \eqref{FFF:TwistReplacement}, exactly reproduces the bosonic twisted lattice \eqref{FFF:T2Z2orbTwist} in the orbifold sector $(h,g)$. Indeed, the triple product identity may be rewritten in the following suggestive form
\begin{equation}
	\left| \vartheta\left[\gamma\atop\delta\right]\,\vartheta\left[\gamma+h\atop\delta+g\right]\,\vartheta\left[1-h\atop 1-g\right]\right|^2= |2\eta^3|^2\,,
	\label{FFF:JacobiTripleProduct}
\end{equation} 
valid whenever the l.h.s. is non-vanishing. In the $(h,g)\neq (0,0)$ sector, out of the four possible values of $\gamma,\delta$, the l.h.s. of \eqref{FFF:JacobiTripleProduct} vanishes in exactly two distinct cases, i.e. when $(\gamma,\delta)=(1,1)$ or $(\gamma,\delta)=(1-h,1-g)$. Therefore, for $(h,g)\neq(0,0)$,
\begin{equation}
	\frac{1}{2}\sum_{\gamma,\delta=0,1} \left| \vartheta\left[\gamma\atop\delta\right]\,\vartheta\left[\gamma+h\atop\delta+g\right]\right|^2 = \left|\frac{2\eta^3}{\vartheta\left[1-h\atop 1-g\right]}\right|^2 \,,
\end{equation}
which indeed reproduces \eqref{FFF:T2Z2orbTwist} for $(h,g)\neq(0,0)$. In a similar fashion, it is possible to establish relations analogous to  \Cref{FFF:TwistReplacement} for more complicated orbifold actions, for instance, involving several $\mathds Z_2$ factors or combinations of rotations and translations.

To be concrete, we shall present here the explicit map between the \ac{FFF} and orbifold formulations of a specific three-generation ${\rm SO}(10)$ model  based on the symmetric $T^6/\mathds{Z}_2\times\mathds{Z}_2$ orbifold model and gauge group ${\rm SO}(10)\times {\rm U}(1)^3\times{\rm SO}(8)^2$, which can be constructed using the symmetric basis of \Cref{FFF:symbasis}. The GGSO matrix of the model, exponentiated in terms of a $\mathds Z_2$-valued matrix $\textbf{G}$ is given by
\begin{equation}
	 C\left[\beta_i\atop\beta_j\right] = (-1)^{G_{ij}}\quad, ~ \textbf{G}=
		\left(\arraycolsep=4pt\begin{array}{c c c c c c c c c c c c}
     1 &  1 &  1 &  1 &  1 &  1 &  1 &  1 &  1 &  1 &  1 & 1\\
     1 &  1 &  1 &  1 &  1 &  1 &  1 &  1 &  1 &  1 &  1 & 1\\
     1 &  1 &  0 &  0 &  0 &  1 &  1 &  0 &  0 &  0 &  0 & 0\\
     1 &  1 &  0 &  0 &  1 &  0 &  1 &  1 &  0 &  0 &  0 & 0\\
     1 &  1 &  0 &  1 &  0 &  0 &  0 &  0 &  0 &  0 &  0 & 1\\
     1 &  1 &  1 &  0 &  0 &  0 &  0 &  0 &  0 &  0 &  1 & 0\\
     1 &  1 &  1 &  1 &  0 &  0 &  0 &  1 &  0 &  0 &  0 & 1\\
     1 &  1 &  0 &  1 &  0 &  0 &  1 &  0 &  0 &  0 &  1 & 0\\
     1 &  0 &  0 &  0 &  0 &  0 &  0 &  0 &  1 &  0 &  0 & 0\\
     1 &  0 &  0 &  0 &  0 &  0 &  0 &  0 &  0 &  1 &  0 & 0\\
     1 &  1 &  0 &  0 &  0 &  1 &  0 &  1 &  0 &  0 &  1 & 1\\
     1 &  1 &  0 &  0 &  1 &  0 &  1 &  0 &  0 &  0 &  1 & 1\\
\end{array}\right) \,.
\label{FFF:GGSOmodelCoeffs}
\end{equation}

This particular vacuum appears in Section 3.2 of \cite{Faraggi:2003yd}. We will map this model to the orbifold framework by explicitly comparing the one-loop partition functions in both representations. This method is based on \cite{florakis:tel-00607408,Florakis:2023aud}, while earlier implementations include \cite{Florakis:2016ani,Florakis:2017ecd,Florakis:2021bws}. We will do this by organizing the boundary condition vectors $\{\beta_i\}$ into a form convenient for the orbifold representation. An inspection of the basis \eqref{FFF:symbasis} clearly identifies the $\mathds Z_2\times \mathds Z_2$ factors with the vectors $\beta_{9}$ and $\beta_{10}$, respectively. We, hence, introduce the notation $(h_1,g_1)$ and $(h_2,g_2)$ to refer to the orbifold parameters labeling the twisted sectors and projection parameters of the $\mathds Z_2$ factors generated by $\beta_{10}$ and $\beta_9$, respectively. We now group together fermions sharing the same boundary conditions, and assign the following boundary condition labels to the fermions appearing in each one of the vectors:
\begin{equation}
	\begin{split}
		&\beta_2 \to (a,b) \,, \\
		&\beta_{2+i} \to (\gamma_i,\delta_i)\quad,\ (i=1,2,\ldots 6) \,,\\		
		&\beta_{1}+\beta_{2}+\sum_{i=1}^{6} \beta_{2+i}+\beta_{11}+\beta_{12} \to (k,\ell)\,, \\
		&\beta_{11}+\beta_{12} \to (\rho,\sigma)\,, \\
	\end{split}
\end{equation}
The logic behind this correspondence is as follows. Before any twist is introduced, the (common) boundary conditions of RNS fermions are denoted $(a,b)$, those of auxiliary fermions realizing the lattice are $(\gamma_i,\delta_i)$, those of fermions realizing the first ${\rm E}_8$ factor are $(k,\ell)$, and those realizing the second ${\rm E}_8$ are $(\rho,\sigma)$. 

The orbifold twists act on top of these assignments and alter the boundary conditions of the fermions associated to $\beta_{9}$, $\beta_{10}$ and $\beta_{12}$. Specifically, the twist in the boundary conditions of the $\mathds Z_2$ orbifold factor associated to $\beta_9$ is labeled by $(h_2,g_2)$, while the corresponding twist due to the $\mathds Z_2$ associated to $\beta_{10}$ is labeled by $(h_1,g_1)$. Finally, $(H,G)$ labels the additional orbifold needed to break the second ${\rm E}_8$ down to a product of ${\rm SO}(8)$ factors. In terms of our assignment, the fermions in $\beta_9$, $\beta_{10}$ and $\beta_{12}$ receive a twist of their boundary conditions as
\begin{equation}
	\begin{split}
		\left. \begin{array}{c}
			(h_2,g_2) \\
			(h_1,g_1)\\
			(H,G)\\
			\end{array}\right\} ~{\rm twists~b.c.~of~fermions~in}~\left\{
			\begin{array}{c}
			\beta_9\\
			\beta_{10}\\
			\beta_{12}\\
			\end{array}\right.
	\end{split}
\end{equation}

To construct the orbifold partition function, it is instructive to start with the ${\rm E}_8\times {\rm E}_8$ heterotic string, compactified on a $T^6$ at the factorized point $T^6=(S^1)^6$ where all radii are equal to $\sqrt{\alpha'/2}$. The partition function of this theory takes the form
\begin{equation}
	Z =\frac{1}{\eta^{12}\bar\eta^{24}}\, \left[\frac{1}{2}\sum_{a,b=0,1}(-1)^{a+b+ab}\vartheta^4\left[a\atop b\right]\right]\Gamma_{6,6}\left[\frac{1}{2}\sum_{k,\ell=0,1}\bar\vartheta^8\left[k\atop\ell\right]\,\frac{1}{2}\sum_{\rho,\sigma=0,1}\bar\vartheta^8\left[\rho\atop\sigma\right]\right]\,.
\end{equation}
In this notation, the RNS fermions $\{\psi^\mu, \chi^{1\ldots 6}\}$ are given the same boundary conditions $(a,b)$ along the two cycles of the worldsheet torus, and these are summed over all spin structures $a,b=0,1$ with the correct spin-statistics (GSO) phase $(-1)^{a+b+ab}$. Note that the additional term $ab$ in the phase is simply a GSO convention. The right-moving Ka{\v c}-Moody fermions $\{\bar\psi^{1\ldots 5}, \bar\eta^1,\bar\eta^2,\bar\eta^3\}$ realize the first ${\rm E}_8$ factor and are similarly given the same boundary conditions $(k,\ell)$, which after appropriate summation over $k,\ell=0,1$ build up the anti-chiral ${\rm E}_8$ lattice. Similarly, $\{\bar\phi^{1\ldots 8}\}$ realize the second ${\rm E}_8$ factor and are assigned the boundary conditions $(\rho,\sigma)$. The Narain lattice of signature $(6,6)$ corresponding to the fermionic point is simply a product of six $(1,1)$ lattices of the type given in eq. \eqref{FFF:lattBoz3}, each one realized by the auxiliary fermions $\{y^i,\omega^i, \bar y^i, \bar\omega^i\}$
\begin{equation}
	\Gamma_{6,6} = \prod_{i=1}^{6}\frac{1}{2}\sum_{\gamma_i,\delta_i=0,1} \vartheta\left[\gamma_i\atop\delta_i\right]\,\bar\vartheta\left[\gamma_i\atop\delta_i\right] \,.
\end{equation} 
This is precisely the theory generated by the basis vectors $\beta_1, \beta_2, \ldots, \beta_8$ together with the combination $\beta_{11}+\beta_{12}$ (the latter simply breaks ${\rm SO}(32)$ to the product of ${\rm E}_8$'s).

Next we consider the orbifold factors. First, notice that including $\beta_{12}$ in our basis\footnote{Or, equivalently, having $\beta_{11}$ and $\beta_{12}$ as separate elements of the basis.}, has the effect of splitting up the boundary conditions of $\bar\phi^{1\ldots 4}$ and $\bar\phi^{5\ldots 8}$ and, hence, may break the second ${\rm E}_8$ down to ${\rm SO}(8)\times{\rm SO}(8)$. In the orbifold representation, we can implement this by defining a $\mathds Z_2$ twist of the boundary conditions of $\bar\phi^{5\ldots 8}$, labelled by $(H,G)$ and by replacing
\begin{equation}
	\bar\vartheta^8\left[\rho\atop\sigma\right] \to (-1)^{HG}\, \bar\vartheta^4\left[\rho\atop\sigma\right]\,\bar\vartheta^4\left[\rho+H\atop\sigma+G\right]\,,
	\label{FFF:E8toSO8square}
\end{equation}
where the phase $(-1)^{HG}$ is required by modularity\footnote{Note that if $(H,G)$ does not couple to the rest of the partition function, summing over $H,G=0,1$ in the r.h.s. of \eqref{FFF:E8toSO8square} reproduces the l.h.s. and ${\rm E}_8$ is regenerated (as it should).}. Similarly, incorporating the $\mathds Z_2\times\mathds Z_2$ orbifold action associated with the elements $\beta_{9}$ and $\beta_{10}$ results in the replacements
\begin{equation}
	\begin{split}
	\vartheta^4\left[a\atop b\right] \to \vartheta\left[a\atop b\right]\,\vartheta\left[a+h_1\atop b+g_1\right]\,\vartheta\left[a+h_2\atop b+g_2\right]\,\vartheta\left[a-h_1-h_2\atop b-g_1-g_2\right]\,, \\
	\bar\vartheta^8\left[k\atop \ell\right] \to \bar\vartheta^5\left[k\atop \ell\right]\,\bar\vartheta\left[k+h_1\atop \ell+g_1\right]\,\bar\vartheta\left[k+h_2\atop \ell+g_2\right]\,\bar\vartheta\left[k-h_1-h_2\atop \ell-g_1-g_2\right]\,, \\
	\end{split}
\end{equation}
for the left and right-moving fermions in the RNS and Ka{\v c}-Moody sectors, respectively, while the lattice fermions are replaced by
\begin{equation}
	\begin{split}
		&\left|\vartheta\left[\gamma_1\atop\delta_1\right]\vartheta\left[\gamma_2\atop\delta_2\right]\right|^2 \to \left|\vartheta\left[\gamma_1\atop \delta_1\right]\vartheta\left[\gamma_1+h_1\atop \delta_1+g_1\right] \vartheta\left[\gamma_2\atop \delta_2\right]\vartheta\left[\gamma_2+h_1\atop \delta_2+g_1\right]\right| \,,\\
		&\left|\vartheta\left[\gamma_3\atop\delta_3\right]\vartheta\left[\gamma_4\atop\delta_4\right]\right|^2 \to \left|\vartheta\left[\gamma_3\atop \delta_3\right]\vartheta\left[\gamma_3+h_2\atop \delta_3+g_2\right] \vartheta\left[\gamma_4\atop \delta_4\right]\vartheta\left[\gamma_4+h_2\atop \delta_4+g_2\right]\right| \,,\\
		&\left|\vartheta\left[\gamma_5\atop\delta_5\right]\vartheta\left[\gamma_6\atop\delta_6\right]\right|^2 \to \left|\vartheta\left[\gamma_5\atop \delta_5\right]\vartheta\left[\gamma_5-h_1-h_2\atop \delta_5-g_1-g_2\right] \vartheta\left[\gamma_6\atop \delta_6\right]\vartheta\left[\gamma_6-h_1-h_2\atop \delta_6-g_1-g_2\right]\right| \,.
	\end{split}
\end{equation}
In terms of boundary condition assignments, factors of the form $|\vartheta|$ should be seen as a short-hand of the real fermion contribution $\vartheta^{1/2}\bar\vartheta^{1/2}$.
The partition function can therefore be cast into the form
\begin{equation}
	\begin{split}
	Z=&\frac{1}{\eta^{12}\bar\eta^{24}}\frac{1}{2^{12}}\sum_{\{A\},\{B\}} (-1)^{a+b+ab+HG+\Phi\left[\{A\}\atop \{B\}\right]}\, \vartheta\left[a\atop b\right]\,\vartheta\left[a+h_1\atop b+g_1\right]\,\vartheta\left[a+h_2\atop b+g_2\right]\,\vartheta\left[a-h_1-h_2\atop b-g_1-g_2\right]  \\
			&\times \left|\vartheta\left[\gamma_1\atop \delta_1\right]\vartheta\left[\gamma_1+h_1\atop \delta_1+g_1\right] \vartheta\left[\gamma_2\atop \delta_2\right]\vartheta\left[\gamma_2+h_1\atop \delta_2+g_1\right]\right| 
			\times \left|\vartheta\left[\gamma_3\atop \delta_3\right]\vartheta\left[\gamma_3+h_2\atop \delta_3+g_2\right] \vartheta\left[\gamma_4\atop \delta_4\right]\vartheta\left[\gamma_4+h_2\atop \delta_4+g_2\right]\right| \\
			&\times \left|\vartheta\left[\gamma_5\atop \delta_5\right]\vartheta\left[\gamma_5-h_1-h_2\atop \delta_5-g_1-g_2\right] \vartheta\left[\gamma_6\atop \delta_6\right]\vartheta\left[\gamma_6-h_1-h_2\atop \delta_6-g_1-g_2\right]\right| \\
			 &\times \bar\vartheta^5\left[k\atop \ell\right]\,\bar\vartheta\left[k+h_1\atop \ell+g_1\right]\,\bar\vartheta\left[k+h_2\atop \ell+g_2\right]\,\bar\vartheta\left[k-h_1-h_2\atop \ell-g_1-g_2\right] 
			 \times\bar\vartheta^4\left[\rho\atop\sigma\right]\,\bar\vartheta^4\left[\rho+H\atop\sigma+G\right]\,,
	\end{split}
	\label{FFF:PartFunctionMap1}
\end{equation}
where $\{A\}=\{a,k,\rho,\{\gamma_i\},h_1,h_2,H\}$ and $\{B\}=\{b,\ell,\sigma,\{\delta_i\},g_1,g_2,G\}$ collectively denote the summation parameters of upper and lower arguments of the theta functions, respectively, while $\Phi$ represents an as-yet unspecified phase that may \emph{a priori} depend on all summation variables in $\{A\}$ and $\{B\}$. Our next step is to constrain this dependence and show how it can be obtained from the \ac{GGSO} coefficients \eqref{FFF:GGSOmodelCoeffs}. In what follows, it will be convenient to assemble the summation parameters into the 12-dimensional column vectors 
\begin{equation}
	\textbf{A}^T=(a,k,\rho,\gamma_1,\ldots,\gamma_6, h_1,h_2,H) \quad,\ \textbf{B}^T=(b,\ell,\sigma,\delta_1,\ldots,\delta_6, g_1,g_2,G) \,.
\end{equation}

It is straightforward to see that $\Phi$ must be invariant under the action of modular transformations on the characteristics. Indeed, using the modular transformation properties of Jacobi theta functions, it can be explicitly checked that the integrand $Z/\tau_2$ of one-loop vacuum amplitude is modular invariant\footnote{Note that the partition function $Z$ defined as in \eqref{FFF:PartFunctionMap1} does not include the non-analytic contribution $(\sqrt{\tau_2})^2=\tau_2^{-1}$, associated to the two transverse non-compact coordinates of $4d$ spacetime. As a result, the modular invariance conditions should actually be imposed on $Z/\tau_2$.} provided $\Phi$ is left invariant (modulo 2) under the action of the modular group on the space of theta characteristics $\{A\}$, $\{B\}$. Specifically, for the $\tau\to\tau+1$ transformation, the relevant action is
\begin{equation}
	\left(\begin{array}{c}
		b\\
		\ell\\
		\sigma\\
		\delta_i\\
		\end{array}\right) \to \left(\begin{array}{c}
		b+a-1\\
		\ell+k-1\\
		\sigma+\rho-1\\
		\delta_i+\gamma_i-1\\
		\end{array}\right)  \quad {\rm and} \quad
	\left(\begin{array}{c}
		g_1\\
		g_2\\
		G\\
	\end{array}\right) \to \left(\begin{array}{c}
		g_1+h_1\\
		g_2+h_2\\
		G+H\\
	\end{array}\right)\,,
	\label{FFF:modCoeff1}
\end{equation}
while for the $\tau\to-1/\tau$ transformation,
\begin{equation}
	\left(\begin{array}{c}
		a\\
		k\\
		\rho\\
		\gamma_i\\
		\end{array}\right) \leftrightarrow \left(\begin{array}{c}
		b\\
		\ell\\
		\sigma\\
		\delta_i\\
		\end{array}\right) \quad {\rm and}\quad
	\left(\begin{array}{c}
		h_1\\
		h_2\\
		H\\
	\end{array}\right) \leftrightarrow \left(\begin{array}{c}
		g_1\\
		g_2\\
		G\\
	\end{array}\right) \,.
	\label{FFF:modCoeff2}
\end{equation}
The exact determination of $\Phi$ specifies the orbifold model and is necessary in order to complete the map from the \ac{FFF}. Indeed, a comparison of eq.\eqref{FFF:PartFunctionMap1} with the general form eq.\eqref{FFF:fpart} of the partition function in the \ac{FFF} (stripped off the $\tau_2$ pre-factors and the modular integral), indicates that the freedom in consistently picking the GGSO coefficients corresponds to the freedom in (consistently) choosing the modular invariant phase $\Phi$. 

To be precise, arrange all theta function factors in \eqref{FFF:PartFunctionMap1} in the same order that the corresponding fermions enter the basis element $\beta_1$ (using complex fermions whenever possible) and define $\textbf{a}$ and $\textbf{b}$ to be the vectors of their upper and lower characteristics, respectively
\begin{equation}
	\textbf{a} = (a, a+h_1, a+h_2, a-h_1-h_2, \ldots)  \quad,\ \textbf{b} = (b,b+g_1, b+g_2, b-g_1-g_2, \ldots)\,.
	\label{FFF:ExtendedBCVs}
\end{equation}
For simplicity, in the above formula we only explicitly display the characteristics of the first few left-moving fermions $\psi^\mu, \chi^{1,2}, \chi^{3,4}$ and $\chi^{5,6}$, but it is clear how to complete the process for all fermions in $\beta_1$, including also the right-moving ones. The end result looks very similar to the boundary condition vectors $\alpha,\beta\in\Xi$ in eq. \eqref{FFF:betas} of the \ac{FFF}, except for the fact that the elements of $\alpha,\beta$ actually correspond to the reduced representatives in the interval $(-1,1]$, i.e. for any fermion $f$ they satisfy\footnote{Of course, the particular example we consider here involves only real fermions, which can be either periodic or anti-periodic so that $\alpha(f),\beta(f)\in\{0,1\}$. However, the method we present here is general and straightforwardly generalizes to rational values as well.} $\alpha(f),\beta(f)\in(-1,1]$. On the other hand, the elements of $\textbf{a},\textbf{b}$ in eq.\eqref{FFF:ExtendedBCVs} do not necessarily lie in $(-1,1]$, but may always brought into this interval by adding suitable even integers. To this end, we denote by $[\textbf{a}]$ the reduced representative of $\textbf{a}$ such that all its elements are in $(-1,1]$. Now, observe, that the set $\{ [\textbf{a}] \}$ of (reduced) boundary condition vectors obtained by allowing the 12 summation parameters $a,k,\rho,\gamma_i, h_1,h_2,H \in \{0,1\}$ to span all allowed values is isomorphic to the subgroup $\Xi$ generated by the 12 basis vectors $\{\beta_i\}$ in the \ac{FFF}, and similarly for $\{ [\textbf{b}]\}$.
Therefore, there exists a bijective map between the \ac{GGSO} coefficients of the \ac{FFF} and the phase $\Phi\left[^\textbf{a}_\textbf{b}\right]$ of the representation \eqref{FFF:PartFunctionMap1}.

In order to determine $\Phi$ in terms of the GSSO coefficients, we express the partition function \eqref{FFF:PartFunctionMap1} in the form
\begin{equation}
	\tau_2\eta^2\bar\eta^2 Z = \frac{1}{2^{12}}\sum_{\textbf{a},\textbf{b}} F\left[\textbf{a}\atop\textbf{b}\right]\, Z\left[\textbf{a}\atop\textbf{b}\right]\,,
	\label{FFF:ZetaThetas3}
\end{equation}
where 
\begin{equation}
	F\left[\textbf{a}\atop\textbf{b}\right]=(-1)^{a+b+ab+HG+\Phi\left[\textbf{a}\atop\textbf{b}\right]}\,,
	\label{FFF:PhaseFactorF}
\end{equation}
contains the phase factor, while $Z\left[\textbf{a}\atop\textbf{b}\right]$ simply contains all left and right moving theta function factors (divided by corresponding Dedekind eta functions) as in \eqref{FFF:ZetaThetas}, arranged in the same order that the fermion boundary conditions appear in $\textbf{a}$ and $\textbf{b}$. In our example the additive group $\Xi$ is a direct sum of 12 $\mathds Z_2$ factors and, hence, $|\Xi|=2^{12}$. The form \eqref{FFF:ZetaThetas3} is very similar to \eqref{FFF:fpart}, except for two important differences. Firstly, \eqref{FFF:fpart} uses the $\Theta$-convention \eqref{FFF:MTheta} for the theta functions instead of the $\vartheta$ one and, secondly, the boundary condition vectors $\alpha,\beta$ in \eqref{FFF:fpart} are of the reduced type (by construction), whereas the $\textbf{a},\textbf{b}$ defined by \eqref{FFF:ExtendedBCVs} are not. The first discrepency is easy to take into account, in view of \eqref{FFF:MTheta}, and amounts to the replacement $\theta\to\Theta$ with the simultaneous inclusion of a phase $\ee^{\frac{\ii\pi}{2}\textbf{a}\cdot\textbf{b}}$. The second discrepancy can be accounted for by noting the identity
\begin{equation}
	\Theta\left[^a_b\right](z;\tau) = \ee^{-\frac{\ii\pi}{2}(a-[a])b}\,\Theta\left[ [a] \atop [b]\right](z;\tau)\,,
\end{equation}
which relates the $\Theta$'s with unreduced characteristics to those of reduced type. In particular, the reduced $\Theta$'s on the r.h.s. are now periodic under both upper and lower arguments. Taking both points into account, \eqref{FFF:ZetaThetas3} can be finally expressed in the form
\begin{equation}
	\tau_2\eta^2\bar\eta^2 Z = \frac{1}{2^{12}}\sum_{\textbf{a},\textbf{b}} C\left[\textbf{a}\atop\textbf{b}\right]\, \hat Z\left[[\textbf{a}]\atop [\textbf{b}]\right]\,,
	\label{FFF:OrbFFFmap}
\end{equation}
with
\begin{equation}
	C\left[\textbf{a}\atop\textbf{b}\right] = F\left[\textbf{a}\atop\textbf{b}\right]\,\ee^{\frac{\ii\pi}{2}\textbf{a}\cdot\textbf{b}-\frac{\ii\pi}{2}(\textbf{a}-[\textbf{a}])\cdot [\textbf{b}]}\,.
	\label{FFF:OrbFFFcoeffs}
\end{equation}
Direct comparison of \eqref{FFF:OrbFFFmap} with \eqref{FFF:fpart} now allows for an exact term-by-term match of the coefficients, once the appropriate change of basis from the $\alpha,\beta \in \Xi$ to the $\textbf{A},\textbf{B}$ is established. In other words, the set of $C\left[\textbf{a}\atop\textbf{b}\right]$'s in the above equation is precisely isomoprhic to the set of \ac{GGSO} coefficients $C\left[\alpha\atop\beta\right]$ of the \ac{FFF}. 

Furthermore, plugging \eqref{FFF:OrbFFFcoeffs} into the modular invariance conditions \eqref{FFF:modCond1}, \eqref{FFF:modCond2} and \eqref{FFF:modCond3c}, we extract the corresponding conditions on the $F$'s that correspond to our basis
\begin{equation}
	\begin{split}
		& F\left[\textbf{a} \atop \textbf{b}-\textbf{a}+1\right] = (-1)^{1+a^2+H^2}\, F\left[\textbf{a}\atop\textbf{b}\right]\,,\\
		&F\left[\textbf{a}\atop\textbf{b}\right] =  F\left[\textbf{b}\atop -\textbf{a}\right]\,,\\
		&F\left[\textbf{a}\atop\textbf{b}+\textbf{b}'\right]=(-1)^a\, F\left[\textbf{a}\atop\textbf{b}\right]\,F\left[\textbf{a}\atop\textbf{b}'\right]\,.
	\end{split}
\end{equation}
In particular, notice that, although the \ac{GGSO} phases $C$ are not symmetric under exchange of upper and lower characteristics, this asymmetry is precisely balanced by the asymmetric phase on the r.h.s of \eqref{FFF:OrbFFFcoeffs}, such that the $F$'s turn out to be symmetric. Since $\Xi$ is a direct sum of $\mathds Z_2$'s, the phases are necessarily real. It is then easy to see that the corresponding conditions on the exponent $\Phi\left[\textbf{a}\atop\textbf{b}\right]$ of the phase appearing in \eqref{FFF:PartFunctionMap1} read
\begin{equation}
	\begin{split}
		& \Phi\left[\textbf{a} \atop \textbf{b}-\textbf{a}+1\right] =\Phi\left[\textbf{a}\atop\textbf{b}\right] \,({\rm mod}\,2)\,,\\
		&\Phi\left[\textbf{a}\atop\textbf{b}\right] = \Phi\left[\textbf{b}\atop -\textbf{a}\right]\, \,({\rm mod}\,2)\,,\\
		&\Phi\left[\textbf{a}\atop\textbf{b}+\textbf{b}'\right]=\Phi\left[\textbf{a}\atop\textbf{b}\right]+\Phi\left[\textbf{a}\atop\textbf{b}'\right]\, \,({\rm mod}\,2)\,.
	\end{split}
	\label{FFF:PhiConditions}
\end{equation}
The first two are the one-loop conditions derived earlier and simply express the fact that $(-1)^\Phi$ is modular invariant, i.e. under \eqref{FFF:modCoeff1} and \eqref{FFF:modCoeff2}. The third one is the two-loop condition arising from factorization and implies that $\Phi$ can be at most linear in the lower characteristics. Together with the second condition, which requires symmetry under the exchange $\textbf{a}\leftrightarrow\textbf{b}$, the phase $\Phi$ is restricted to be of the form
\begin{equation}
	\Phi\left[\textbf{a}\atop\textbf{b}\right] = \textbf{A}^T\textbf{M}\textbf{B} = \sum_{i,j=1}^{12} A_i M_{ij} B_j\,,
	\label{FFF:ModPhaseFerm}
\end{equation}
where $\textbf{M}$ is a constant $12\times 12$ symmetric matrix (defined modulo 2), which is in one-to-one correspondence with the \ac{GGSO} matrix \eqref{FFF:GGSOmodelCoeffs}. The matrix $\textbf{M}$ is further constrained by the first condition in \eqref{FFF:PhiConditions}, which implies $M_{ii}=\sum_{j=1}^9 M_{ij}$ for all $i=1,\ldots, 12$. However, not all those conditions are independent. Their sum trivially vanishes (modulo 2), since it involves adding the upper and lower triangular parts which, however, are equal due to the symmetry of $\textbf{M}$. In general, for $n$ basis vectors, there are $\frac{1}{2}n(n-1)$ conditions from the symmetry of $M_{ij}$ but only $(n-1)$ conditions from the modular $T$-transformation. As a result, $M_{ij}$ has $\frac{1}{2}n(n-1)+1$ independent elements, which is consistent with the fact that there are $2^{\frac{n(n-1)}{2}+1}$ fermionic models with the given basis.

We will now establish the precise relation between the 12-dimensional matrix of \ac{GGSO} coefficients $C\left[\beta_i\atop\beta_j\right]$ and $\textbf{M}$. 
Until we have to deal with the asymmetric phase on the r.h.s. of \eqref{FFF:OrbFFFcoeffs}, we can work modulo 2 and effectively identify $\textbf{a}$ with $[\textbf{a}]$. Take an arbitrary element $\alpha =\textbf{a}\in \Xi$ and expand it in the $\{\beta_i\}$ basis of the \ac{FFF} as $\textbf{a}=\lambda_i(\textbf{a}) \beta_i$. The components $\lambda_i(\textbf{a})$ will clearly be linear combinations of the $A_i$'s, namely $\lambda_i(\textbf{a}) = \tilde S_{ij} A_j(\textbf{a})$ for some invertible matrix $\tilde S_{ij}$ independent of $\textbf{a}$ which one may easily identify. We can then interpret $A_i(\textbf{a})$ as the linear functions 
\begin{equation}
	A_i(\textbf{a}) = (\tilde S^{-1})_{ij} \beta^\star_{j}(\textbf{a})\,,
\end{equation}
in terms of the dual basis $\{\beta^\star_j\}$. In particular, $A_j(\beta_i) =(\tilde S^{-T})_{ij} \equiv S_{ij}$. Having determined the matrix encoding the change of basis $S_{ij}$ allows all parameters $a,b,k,\ell,\ldots$ to be uniquely fixed in terms of the basis vectors $\beta_i,\beta_j$ appearing as upper and lower characteristics of $C\left[\beta_i\atop\beta_j\right]$. For example, $a\to A_1(\beta_i)=S_{i1}$ while $b\to B_1(\beta_j)=S_{j2}$, and similarly for the others.
In this notation, we have
\begin{equation}
	\Phi\left[\beta_i\atop\beta_j\right] = \sum_{r,s=1}^{12}A_r(\beta_i)M_{rs}B_s(\beta_j) = (\textbf{S}\textbf{M}\textbf{S}^T)_{ij}\,.
\end{equation}
Now define the $\mathds Z_2$-valued \ac{GGSO} exponent matrix $G_{ij}\in\{0,1\}$ via $C\left[\beta_i\atop\beta_j\right]= (-1)^{G_{ij}}$ and assemble the exponents (modulo 2) of the phase factors of \eqref{FFF:PhaseFactorF} and \eqref{FFF:OrbFFFcoeffs} into
\begin{equation}
	L\left[^\textbf{a}_\textbf{b}\right] = a+b+ab+HG-\frac{1}{2}(\textbf{a}-[\textbf{a}])\cdot\textbf{b}\,.
\end{equation}
The evaluation of $L_{ij}\equiv L\left[\beta_i\atop\beta_j\right]$ should be performed with special care, since modulo 2 periodicities \emph{do} matter in the last term.
The matrix $\textbf{M}$ is then obtained (modulo 2) by
\begin{equation}
	\textbf{M} = \textbf{S}^{-1} (\textbf{G}+\textbf{L})\textbf{S}^{-T} \,.
\end{equation}
Carrying out these steps for the example model at hand, we obtain
\begin{equation}
	\textbf{M} = \left(
\begin{array}{cccccccccccc}
 0 & 0 & 0 & 0 & 0 & 0 & 0 & 0 & 0 & 0 & 0 & 0 \\
 0 & 0 & 0 & 0 & 1 & 0 & 0 & 0 & 1 & 0 & 0 & 0 \\
 0 & 0 & 0 & 0 & 0 & 1 & 1 & 1 & 1 & 0 & 0 & 0 \\
 0 & 0 & 0 & 0 & 0 & 0 & 1 & 1 & 0 & 0 & 0 & 0 \\
 0 & 1 & 0 & 0 & 0 & 1 & 0 & 1 & 1 & 1 & 1 & 0 \\
 0 & 0 & 1 & 0 & 1 & 0 & 0 & 0 & 0 & 0 & 0 & 1 \\
 0 & 0 & 1 & 1 & 0 & 0 & 0 & 0 & 0 & 0 & 0 & 0 \\
 0 & 0 & 1 & 1 & 1 & 0 & 0 & 0 & 1 & 0 & 0 & 1 \\
 0 & 1 & 1 & 0 & 1 & 0 & 0 & 1 & 0 & 1 & 1 & 0 \\
 0 & 0 & 0 & 0 & 1 & 0 & 0 & 0 & 1 & 0 & 0 & 0 \\
 0 & 0 & 0 & 0 & 1 & 0 & 0 & 0 & 1 & 0 & 0 & 0 \\
 0 & 0 & 0 & 0 & 0 & 1 & 0 & 1 & 0 & 0 & 0 & 0 \\
\end{array}
\right)\,.
\end{equation}
Plugging this matrix into \eqref{FFF:ModPhaseFerm}, it is straightforward to extract the modular invariant phase $\Phi$ in terms of the summation parameters. The partition function \eqref{FFF:PartFunctionMap1} with $\Phi$ determined using this procedure then precisely reproduces eq. \eqref{FFF:fpart} term by term. 

Up to this point, the map was clearly a bijective one. However, the partition function in \Cref{FFF:PartFunctionMap1} is \emph{not yet} in the orbifold representation. After all, all we did so far was re-organize the terms of \eqref{FFF:fpart} into a useful intermediate form. The actual orbifold representation, however, is just around the corner. Notice that the orbifold partition function is not formulated in terms of $\gamma_i,\delta_i$ but should be expressed in terms of Narain lattices with twists and shifts, at the special loci in moduli space compatible with bosonization. 

To this end, it is necessary to obtain a generalized version of \eqref{FFF:T2Z2orbTwist} and \eqref{FFF:TwistReplacement} applicable to our case, where all six compactified coordinates are fermionized with different boundary conditions $(\gamma_i,\delta_i)$. The contribution of the fermionized coordinates will be organized as a product of three lattices of signature $(2,2)$, which will now contain both shifts and twists. For example, for the first lattice, we have
\begin{equation}
	\begin{split}
	\Gamma_{2,2}\left[ \begin{array}{c c|c} H_1 & H_2 & h_1 \\ G_1 & G_2 & g_1\end{array} \right](2\ii\alpha', \ii) = \frac{1}{4}\sum_{\gamma_1,\delta_1=0,1\atop \gamma_2,\delta_2=0,1} & (-1)^{\gamma_1 G_1+\delta_1 H_1+H_1G_1}\, (-1)^{\gamma_2 G_2+\delta_2 H_2+H_2 G_2} \\
		&\times \left|\vartheta\left[\gamma_1\atop \delta_1\right]\vartheta\left[\gamma_1+h_1\atop \delta_1+g_1\right] \vartheta\left[\gamma_2\atop \delta_2\right]\vartheta\left[\gamma_2+h_1\atop \delta_2+g_1\right]\right| \,,
	\end{split}
	\label{FFF:TwistedShiftedLatticeFerm}
\end{equation}
and similarly for the remaining two lattices. Here, $(h_1,g_1)$ are again associated to the $\mathds Z_2$ twist that rotates the coordinates of the first 2-torus. The new parameters $(H_1,G_1)$ and $(H_2,G_2)$ are similarly ascribed to additional $\mathds Z_2$ orbifold factors which, respectively, act as translations (shifts) along the two cycles of the $T^2$. The fermionization point now corresponds\footnote{The factorization point may appear to be in conflict with the value $T/\alpha'=U=\ii$ obtained earlier in eq. \eqref{FFF:T2Z2orbTwist}. However, the difference is that the lattice of eq. \eqref{FFF:TwistedShiftedLatticeFerm} contains two independent shifts acting along the two $T^2$ directions and the factorization point is also sensitive to their embedding in the space of momenta and windings. The factorization point $T/2\alpha'=U=\ii$ corresponds to both translations acting as momentum shifts.} to $T=2\ii\alpha'$, $U=\ii$, while for vanishing twist $h_1=g_1=0$ one recovers the shifted lattice
\begin{equation}
	\Gamma_{2,2}\left[\begin{array}{cc|c} H_1 & H_2 & 0\\G_1 & G_2 & 0\\ \end{array}\right](T,U) = \sum_{m_i,n_i\in\mathds Z} (-1)^{m_1 G_1+m_2 G_2} \,q^{\frac{\alpha'}{4}|P_L|^2}\,\bar q^{\frac{\alpha'}{4}|P_R|^2}\,,
\end{equation}
where now the complexified momenta also depend on the shift parameters
\begin{equation}
	\begin{split}
		&P_L = \frac{m_2-Um_1+T\left(n_1+\frac{H_1}{2}+U\left(n_2+\frac{H_2}{2}\right)\right)}{\sqrt{T_2 U_2}} \,,\\ 
		&P_R = \frac{m_2-Um_1+\bar T\left(n_1+\frac{H_1}{2}+U\left(n_2+\frac{H_2}{2}\right)\right)}{\sqrt{T_2 U_2}}\,.
	\end{split}
	\label{FFF:complexMomenta2}
\end{equation}
It is easy to invert \eqref{FFF:TwistedShiftedLatticeFerm}  and rewrite it in a more convenient form
\begin{equation}
	\begin{split}
		\frac{1}{4}\sum_{\gamma_1,\delta_1=0,1\atop\gamma_2,\delta_2=0,1}& (-1)^{\gamma_1 Y_1+\delta_1 X_1+\gamma_2 Y_2+\delta_2 X_2}
			\,\left|\vartheta\left[\gamma_1\atop \delta_1\right]\vartheta\left[\gamma_1+h_1\atop \delta_1+g_1\right] \vartheta\left[\gamma_2\atop \delta_2\right]\vartheta\left[\gamma_2+h_1\atop \delta_2+g_1\right]\right| \\
			 = \frac{1}{2^2} &\sum_{H_1,G_1=0,1\atop H_2,G_2=0,1} (-1)^{H_1 Y_1+G_1 X_1+X_1 Y_1}\,(-1)^{H_2 Y_2+G_2 X_2+X_2 Y_2} \, \Gamma_{2,2}\left[ \begin{array}{c c|c} H_1 & H_2 & h_1 \\ G_1 & G_2 & g_1\end{array} \right](2\ii\alpha', \ii)\,,
	\end{split}
	\label{FFF:TwistedShiftedLatticeFerm2}
\end{equation}
valid for any $\mathds Z_2$-valued parameters $X_1, X_2, Y_1, Y_2$. This may be now used in order to replace the $(\gamma_i,\delta_i)$-coupled theta functions with Narain lattices. Doing so, the partition function of the theory can finally be brought into its orbifold representation
\begin{equation}
	\begin{split}
	Z=&\frac{1}{\eta^{12}\bar\eta^{24}}\frac{1}{2^{12}}\sum_{\{A'\},\{B'\}} (-1)^{a+b+ab+HG+\Phi' \left[\{A'\}\atop \{B'\}\right]}\, \\
			& \times \vartheta\left[a\atop b\right]\,\vartheta\left[a+h_1\atop b+g_1\right]\,\vartheta\left[a+h_2\atop b+g_2\right]\,\vartheta\left[a-h_1-h_2\atop b-g_1-g_2\right]  \\
			& \times \Gamma_{2,2}\left[\begin{array}{cc|c} H_1 & H_2 & h_1\\G_1 & G_2 & g_1\\ \end{array}\right] \,
			  \Gamma_{2,2}\left[\begin{array}{cc|c} H_3 & H_4 & h_2\\G_3 & G_4 & g_2\\ \end{array}\right] \,
			\Gamma_{2,2}\left[\begin{array}{cc|c} H_5 & H_6 & h_1+h_2\\G_5 & G_6 & g_1+g_2\\ \end{array}\right] \\
			 &\times \bar\vartheta^5\left[k\atop \ell\right]\,\bar\vartheta\left[k+h_1\atop \ell+g_1\right]\,\bar\vartheta\left[k+h_2\atop \ell+g_2\right]\,\bar\vartheta\left[k-h_1-h_2\atop \ell-g_1-g_2\right] 
			 \times\bar\vartheta^4\left[\rho\atop\sigma\right]\,\bar\vartheta^4\left[\rho+H\atop\sigma+G\right]\,,
	\end{split}
	\label{FFF:PartFunctionMap2}
\end{equation}
where $\{A'\}=\{a,k,\rho, \{ H_i\},h_1,h_2,H\}$ and $\{B'\}=\{b,\ell,\sigma,\{G_i\},g_1,g_2,G\}$ and $i=1,\ldots, 6$ are the new summation variables. Importantly, $\Phi'$ is the new modular invariant phase obtained from $\Phi$ under the phase substitution implied by \eqref{FFF:TwistedShiftedLatticeFerm2}.
Concretely, one finds
\begin{equation}
	\begin{split}
		&\Phi' = k\ell +[kg_1+\ell h_1+h_1 g_1]+[kg_2+\ell h_2+h_2 g_2]\\
			&+ [G_1 \rho+ H_1\sigma+H_1 G_1] + [G_2(\rho+H)+H_2(\sigma+G)+H_2 G_2] \\
			&+ [G_3(k+H)+H_3(\ell+G)+H_3 G_3]+[G_4(k+\rho +H)+ H_4(\ell+\sigma+G)] \\
			&+ [G_5(k+H)+H_5(\ell+G)+H_5 G_5]+ [G_6(k+\rho+H)+H_6(\ell+\sigma+G)]\\
			&+[h_1 g_2+g_1 h_2] + [h_1(G_3+G_4+G_5+G_6)+g_1(H_3+H_4+H_5+H_6)]\\
			& +[h_2(G_3+G_4+G_5+G_6)+g_2(H_3+H_4+H_5+H_6)]\\
			&+ (14) + (23) + (34)+ (46) +(35)+(56)\,,
	\end{split}
	\label{FFF:explicitPhase}
\end{equation}
where in the last line, the terms of the form $(ij)$ stand for $G_i H_j+H_i G_j$, and we have grouped terms which are separately modular invariant into square brackets. In this representation it is possible to recognize the orbifold action. Aside from the familiar $\mathds Z_2\times \mathds Z_2$ rotating the first and second pair of 2-tori respectively, the second, third and fourth lines of \eqref{FFF:explicitPhase} indicate the free actions $(-1)^{F_2}\sigma_1$, $(-1)^{F_2}\sigma_2$, $(-1)^{F_1}\sigma_3$, $(-1)^{F_1+F_2}\sigma_4$, $(-1)^{F_1}\sigma_5$ and $(-1)^{F_1+F_2}\sigma_6$, where $\sigma_I$ denote the order-2 shifts along each direction and $F_1$, $F_2$ are the ``fermion numbers'' associated to the spinorial representations of the original $\text{E}_8$ factors, respectively. The remaining terms in the phases correspond to chirality conventions and discrete torsion phases.

As a cross-check, it is instructive to compare the contributions of bosonic and fermionic states to the partition function in both the \ac{FFF} and in the orbifold formulations using \eqref{FFF:fpart} and \eqref{FFF:PartFunctionMap2}, respectively, at the fermionic point. One obtains
\begin{align}
Z_{\rm B}=-Z_{\rm F}&= 2\overline{q}^{-1}+872+120 q^{1/2} \overline{q}^{-1/2}
+ 32 q \overline{q}^{-1}+4080 q^{1/8}\overline{q}^{1/8}\nonumber\\
&\ \ \ \ +16 q^{9/8}\overline{q}^{-7/8} + 704 q^{5/8}\overline{q}^{-3/8}
+\dots \ ,
\end{align}
and the contributions match, as they should. Note that the contribution of bosonic states $Z_{\rm B}$ exactly cancels that of the fermionic ones $Z_{\rm F}$ due to the unbroken spacetime supersymmetry of the theory. Note, furthermore, that each term comes with positive integer multiplicity, as required for a correct particle interpretation.

With the map from the \ac{FFF} to the orbifold representation established, it is possible to reinstate the moduli dependence back into the Narain lattices and deform the theory away from the fermionic point. In particular, this is the natural path to take if one is interested in studying supersymmetry breaking in \ac{FFF} models. For instance, in the context of the particular vacuum discussed here, it is easy to see that $\mathcal N=1$ supersymmetry is unbroken, as may be checked either by explicitly constructing the supercharges or by means of Jacobi identities. In non-supersymmetric constructions, one may extract and study the gravitino mass $m_{3/2}$ as a function of the compactification moduli, and obtain conditions for the breaking to be spontaneous directly in terms of the \ac{GGSO} coefficients\footnote{For a discussion, see \cite{Florakis:2016ani}.}.

It is also important to mention that the orbifold representation in terms of the phase \eqref{FFF:explicitPhase} is not unique. There are equivalent representations related by lattice redefinitions and T-dualities, arising from the different ways of performing the summations over $\gamma_i,\delta_i$. We shall not elaborate on this here, but instead refer the interested reader to \cite{Florakis:2023aud} for a detailed discussion.


\bibliographystyle{alpha}

\newcommand{\etalchar}[1]{$^{#1}$}


\begin{acronym}
  \acro{BBN}{big bang nucleosynthesis}
  \acro{BSM}{beyond the standard model}
  \acro{CFT}{conformal field theory}
  \acro{CY}{Calabi--Yau}
  \acro{EFT}{effective field theory}
  \acro{FCNC}{flavor changing neutral current}
  \acro{FFF}{Free Fermionic Formulation}
  \acro{FI}{Fayet--Iliopoulos \cite{Fayet:1974jb}} 
  \acro{GGSO}{Generalized GSO Projections}
  \acro{GS}{Green--Schwarz \cite{Green:1984sg}}
  \acro{GSO}{Gliozzi--Scherk--Olive}
  \acro{GUT}{Grand Unified Theory}
  \acro{KK}{Kaluza--Klein}
  \acro{LHC}{Large Hadron Collider}
  \acro{OPE}{operator product expansion}
  \acro{MSSM}{minimal supersymmetric standard model}
  \acro{PS}{Pati--Salam \cite{Pati:1974yy}} 
  \acro{QFT}{quantum field theory}
  \acro{SB}{symmetry based}
  \acro{SM}{standard model}
  \acro{SUSY}{supersymmetry}
  \acro{UV}{ultraviolet}
  \acro{VEV}{vacuum expectation value}
\end{acronym}

\end{document}